\pdfoutput=1
\documentclass[%
  aps,
  prb,
  showpacs,
  citeautoscript,
  superscriptaddress,
  amsmath,
  amssymb,
  reprint
]{revtex4-1}

\usepackage[utf8]{inputenc}
\usepackage[T1]{fontenc}
\usepackage[colorlinks=true,urlcolor=blue,linkcolor=blue,citecolor=blue]{hyperref}

\usepackage{graphicx}
\usepackage[caption=false]{subfig}
\usepackage{dcolumn}
\usepackage{multirow}
\usepackage{mhchem}
\usepackage{braket}

\newcommand\textlcsc[1]{\textsc{\MakeLowercase{#1}}}

\newcommand{\M}[1]{{\bf #1}}
\newcommand{\V}[1]{\boldsymbol #1}
\newcommand{\dee}{\text{d}}
\newcommand{\eh}{\text{eh}}
\renewcommand{\Im}{\text{Im}}
\newcommand{\ie}{\emph{i.e.}}
\newcommand{\eg}{\emph{e.g.}}
\newcommand{\etal}{\emph{et al.}}

\begin{document}

\title{Kohn--Sham decomposition in real-time time-dependent density-functional theory:\\
An efficient tool for analyzing plasmonic excitations
}

\date{\today}

\author{Tuomas\ P.\ Rossi}
\email{tuomas.rossi@alumni.aalto.fi}
\affiliation{COMP Centre of Excellence, Department of Applied Physics,
Aalto University School of Science, Espoo, Finland}

\author{Mikael\ Kuisma}
\email{mikael.j.kuisma@jyu.fi}
\affiliation{Department of Physics, Chalmers University of Technology, Gothenburg, Sweden}
\affiliation{Department of Chemistry, Nanoscience Center,
University of Jyv\"askyl\"a, Jyv\"askyl\"a, Finland}

\author{Martti\ J.\ Puska}
\affiliation{COMP Centre of Excellence, Department of Applied Physics,
Aalto University School of Science, Espoo, Finland}

\author{Risto\ M.\ Nieminen}
\affiliation{COMP Centre of Excellence, Department of Applied Physics,
Aalto University School of Science, Espoo, Finland}

\author{Paul\ Erhart}
\email{erhart@chalmers.se}
\affiliation{Department of Physics, Chalmers University of Technology, Gothenburg, Sweden}

\begin{abstract}
The real-time-propagation formulation of time-dependent density-functional
theory (RT-TDDFT) is an efficient method for modeling the optical response of
molecules and nanoparticles. Compared to the widely adopted linear-response
TDDFT approaches based on, \emph{e.g.}, the Casida equations, RT-TDDFT appears,
however, lacking efficient analysis methods. This applies in particular to a
decomposition of the response in the basis of the underlying single-electron
states. In this work, we overcome this limitation by developing an analysis
method for obtaining the Kohn--Sham electron-hole decomposition in RT-TDDFT.
We demonstrate the equivalence between the developed method and the Casida
approach by a benchmark on small benzene derivatives. Then, we use the method
for analyzing the plasmonic response of icosahedral silver nanoparticles up to
Ag$_{561}$. Based on the analysis, we conclude that in small nanoparticles
individual single-electron transitions can split the plasmon into multiple
resonances due to strong single-electron--plasmon coupling whereas in larger
nanoparticles a distinct plasmon resonance is formed.
\end{abstract}


\maketitle

\section{Introduction}
\label{sec:introduction}

Time-dependent density-functional theory (TDDFT) \cite{Runge1984} built on top
of Kohn--Sham (KS) density-functional theory (DFT) \cite{Hohenberg1964,Kohn1965}
is a powerful tool in computational physics and chemistry for accessing the
optical properties of matter. \cite{Marques2012,Ullrich2012} Starting from
seminal works on jellium nanoparticles, \cite{Ekardt1984, Puska1985, Beck1987}
TDDFT has become a standard tool for modeling plasmonic response from a
quantum-mechanical perspective, \cite{Morton2011, Varas2016} and proven to be
useful for calculating the response of individual nanoparticles,
\cite{Prodan2003Electronic, Aikens2008, Zuloaga2010, Weissker2011, Li2013, Piccini2013, Burgess2014, Barcaro2014, Weissker2015, Bae2015, Zapata2016}
and their compounds
\cite{Zuloaga2009, Song2011, Song2012, Marinica2012, Zhang2014, Varas2015, Barbry2015, Kulkarni2015, Rossi2015Quantized, Marchesin2016, Lahtinen2016}
as well as other plasmonic materials.
\cite{Manjavacas2013, Andersen2013Plasmons, Andersen2014, Lauchner2015}
Additionally, a number of models and concepts have been developed for
quantifying and understanding plasmonic character within the TDDFT framework.
\cite{Gao2005, Yan2007, Bernadotte2013, Malola2013, Guidez2012, Guidez2014, Yasuike2011, Casanova2016, Townsend2012, Townsend2014, Ma2015, Bursi2016}
Thus, in conjunction with other theoretical and computational methods
\cite{Esteban2012, Stella2013, Chen2015, Yan2015, Teperik2016, Ciraci2016, David2016, Christensen2017}
and experimental developments,
\cite{Ciraci2012, Scholl2012, Haberland2013, Savage2012, Banik2012, Banik2013, Scholl2013, Tan2014, Zhang2016, Sanders2016, Mertens2016}
TDDFT is a valuable tool for understanding quantum effects within the
nanoplasmonics field. \cite{Tame2013, Zhu2016}
Recent methodological advances and a steady increase in computational power
have extended the system size that can be treated at the TDDFT level, enabling
the computational modeling of plasmonic phenomena in noble metal nanoparticles
of several nanometers in diameter.
\cite{Iida2014, Kuisma2015, Baseggio2015, Baseggio2016, Koval2016}

TDDFT in the linear-response regime is usually formulated in frequency space
\cite{Casida1995, Petersilka1996}
in terms of the Casida matrix expressed in the Kohn--Sham electron-hole space.
\cite{Casida1995, Casida2009}
The calculations are commonly performed by diagonalizing the Casida matrix
directly or by solving the equivalent problem with different iterative subspace
algorithms.
\cite{Bauernschmitt1996, Stratmann1998, Walker2006, Andrade2007}
The real-time-propagation formulation of TDDFT (RT-TDDFT)
\cite{Yabana1996,Yabana2006}
is a computationally efficient alternative to frequency-space approaches with
favorable scaling with respect to system size, \cite{Sander2017} and has the
additional advantage of being also applicable to the non-linear regime.
However, RT-TDDFT results are often limited to absorption spectra or to
analyses of transition densities, apart from a few exceptions focusing on
characterizing plasmonic \cite{Townsend2012, Townsend2014, Ma2015, Yan2016} or
other electronic excitations. \cite{Li2011, Li2015, Repisky2015, Kolesov2016}
In contrast, the Casida approach directly enables an extensive analysis in
terms of the KS electron-hole decomposition of the excitations and thereby
readily yields quantum-mechanical understanding of the plasmonic response.
\cite{Yasuike2011, Guidez2012, Guidez2014, Bernadotte2013, Malola2013, Malola2014, Malola2015, Baseggio2015, Baseggio2016, Casanova2016}

In this work, we remedy the lack of analysis tools in RT-TDDFT and demonstrate
that the decomposition of the electronic excitations in terms of the underlying
KS electron-hole space can be obtained within RT-TDDFT, in equivalent fashion
to the Casida approach. We have combined the analysis method with a recent
RT-TDDFT implementation \cite{Kuisma2015} based on the linear combination of
atomic orbitals (LCAO) method \cite{Larsen2009} that is part of the open source
\textlcsc{GPAW} code. \cite{Mortensen2005, Enkovaara2010, GPAW}

By using the developed method, we perform a KS decomposition analysis of the
plasmon formation in a series of icosahedral silver nanoparticles comprising
\ce{Ag55}, \ce{Ag147}, \ce{Ag309}, and \ce{Ag561}. We observe that while in
\ce{Ag147} and larger nanoparticles a distinct plasmon resonance is formed from
the superposition of single-electron transitions, in the small \ce{Ag55}
nanoparticle individual single-electron transitions still have a strong effect
on the plasmonic response and cause the splitting of the plasmon resonance.

The structure of the article is as follows. In Sec.~\ref{sec:methods} we derive
the linear response of the time-dependent density matrix in the KS
electron-hole space. We review the formulation of the same quantity in the
Casida approach and describe the decomposition of the photo-absorption spectrum
in KS electron-hole contributions. In Sec.~\ref{sec:results} we benchmark the
numerical accuracy of the implemented method by analyzing the KS decomposition of small
benzene derivatives using both the real-time-propagation and the Casida method.
This is followed by an analysis of the plasmonic response of large silver
nanoparticles, which yields microscopic insight into the plasmon formation in
nanoparticles. In Sec.~\ref{sec:discussion} we discuss the general features of
the presented methodology. Our work is concluded in Sec.~\ref{sec:conclusions}.

\section{Methods}
\label{sec:methods}

\subsection{Linear response of the density matrix in the real-time propagation method}
\label{sec:rttdft}

The time-dependent Kohn--Sham equation is defined as
\begin{equation}
    i \frac{\partial}{\partial t} \psi_n(\V r,t) = H_{\text{KS}}(t) \psi_n(\V r,t) ,
    \label{eq:tdks}
\end{equation}
where $H_{\text{KS}}(t)$ is the time-dependent KS Hamiltonian and
$\psi_n(\V r,t)$ is a KS wave function.
The density matrix operator is defined as
\begin{equation}
    \rho(t) = \sum_n \ket{\psi_n(t)} f_n \bra{\psi_n(t)} ,
\end{equation}
where $f_n$ is an occupation factor of the $n$th KS state.
In order to proceed with KS decomposition, we express the density matrix
in the KS basis, spanned by the ground-state KS orbitals $\psi_n^{(0)}(\V r)$,
which fulfill the ground-state KS equation
\begin{equation}
    H_{\text{KS}}^{(0)} \psi_n^{(0)}(\V r) = \epsilon_n \psi_n^{(0)}(\V r) ,
    \label{eq:gs}
\end{equation}
where $H_{\text{KS}}^{(0)}$ is the ground-state KS Hamiltonian and $\epsilon_n$
the KS eigenvalue of $n$th state.
The density matrix can be written in this KS basis as
\begin{align}
    \rho_{nn'}(t) &= \braket{\psi^{(0)}_{n} | \rho(t) | \psi^{(0)}_{n'}} \nonumber \\
    &= \sum_m \braket{\psi^{(0)}_{n} | \psi_m(t)} f_m \braket{\psi_m(t) | \psi^{(0)}_{n'}} .
    \label{eq:rhonn-braket}
\end{align}
This equation establishes a link between a time-dependent density matrix and
the usual KS (electron-hole) basis set used in linear-response calculations,
see Sec.~\ref{sec:casida}. Previously, similar or related quantities have been
used within the real-time propagation method for analysing the response.
\cite{Li2011, Li2015, Townsend2012, Townsend2014, Ma2015, Repisky2015, Yan2016}

When the real-time propagation method is applied in the linear-response regime,
the usual approach is to use a $\delta$-pulse perturbation.
\cite{Yabana1996, Yabana2006}
This corresponds to the Hamiltonian
\begin{equation}
    H_{\text{KS}}(t) = H^{(0)}_{\text{KS}} + z K_z \delta(t) ,
    \label{eq:kickham}
\end{equation}
where the interaction with external electromagnetic radiation is taken within
the dipole approximation. The electric field is assumed to be aligned along
the $z$ direction and the constant $K_z$ is proportional to the external
electric field strength, which is assumed to be small enough to induce only
negligible non-linear effects. After the perturbation by the $\delta$-pulse at
$t=0$, Eq.~\eqref{eq:tdks} is propagated in time and the quantities of interest
are recorded during the propagation. As a post-processing step time-domain
quantities, \eg, $\rho_{nn'}(t)$, can be Fourier transformed into the frequency
domain.

It is important to note that the size of the density matrix $\rho_{nn'}(t)$ can
be significantly reduced since only its electron-hole part is required in
linear-response theory. \cite{Casida1995, Casida2009} It is thus sufficient to
consider only $\rho_{ia}(t)$, where $i$ and $a$ represent occupied and
unoccupied KS states, respectively. Then, we obtain the linear-response of the
density matrix in electron-hole space as
\begin{equation}
    \delta\rho^{z}_{ia}(\omega) = \frac{1}{K_z} \int_{0}^{\infty}
    \left[ \rho^{z}_{ia}(t) - \rho_{ia}(0^-) \right] e^{i\omega t} \dee t + \mathcal{O}(K_z),
    \label{eq:rho td}
\end{equation}
where $\rho_{ia}(0^-)$ is the initial density matrix before the $\delta$-pulse
perturbation and the superscript $z$ indicates the direction of the
perturbation.

In common TDDFT implementations, there is no mechanism for energy dissipation
and the lifetime of excitations is infinite. A customary way to restore a
finite lifetime is to apply the substitution $\omega \to \omega + i \eta$,
where the parameter $\eta$ is small. This leads to an exponentially decaying
term in the integrand in Eq.~\eqref{eq:rho td}, \ie,
$e^{i\omega t} \to e^{i\omega t} e^{-\eta t}$,
and to the Lorentzian line shapes in the frequency domain. The decaying
integrand also means that a finite propagation time is sufficient in practical
calculations. The Gaussian line shapes can be obtained by replacing the
Lorentzian decay $e^{-\eta t}$ with the Gaussian decay function
$e^{- (\sigma t)^2/2}$, where the parameter $\sigma$ determines the spectral
line width.

\subsubsection*{Implementation}

We have implemented the density matrix formalism outlined above in the RT-TDDFT
code \cite{Kuisma2015} that is part of the open source \textlcsc{GPAW} package
\cite{Mortensen2005,Enkovaara2010,GPAW}. Our implementation uses the LCAO
basis set\cite{Larsen2009} and the projector-augmented wave (PAW)
\cite{Blochl1994} method. In the LCAO method the wave function
$\psi_n(\V r, t)$
is expanded in localized basis functions $\phi_\mu(\V r)$ centered at atomic
coordinates
\begin{equation}
    \psi_n(\V r, t) = \sum_\mu \phi_\mu(\V r) C_{\mu n}(t)
    \label{eq:lcaowf}
\end{equation}
with expansion coefficients $C_{\mu n}(t)$.
The density matrix is reads in the LCAO basis set as
\begin{equation}
    \rho_{\mu\nu}(t) = \sum_n C_{\mu n}(t) f_n C_{\nu n}^*(t) .
    \label{eq:rhomm}
\end{equation}
Then, Eq.~\eqref{eq:rhonn-braket} can be written in LCAO formalism as (using
implied summation over repeated indices)
\begin{equation}
    \rho_{nn'}(t) = C^{(0)*}_{\mu n} S_{\mu \mu'} \rho_{\mu'\nu'}(t) S^*_{\nu \nu'} C^{(0)}_{\nu n'} ,
    \label{eq:rhonn}
\end{equation}
where
$S_{\mu\mu'} = \int \phi^*_{\mu}(\V r) \phi_{\mu'}(\V r) \dee \V r$
is the overlap integral of the basis functions.
A detailed derivation of Eq.~\eqref{eq:rhonn} is given in Appendix, in which it
is shown that the PAW transformation affects only the evaluation of the overlap
integral.

The emphasis in our implementation is to minimize the computational footprint
of the analysis methods. Thus, instead of calculating Eq.~\eqref{eq:rhonn} at
every time step during the time propagation, we only store the
already-calculated matrix $C^z_{\mu n}(t)$ at every time step. Then, as a
post-processing step, we calculate $\rho^z_{\mu\nu}(t)$ with
Eq.~\eqref{eq:rhomm} and Fourier transform the result to obtain
$\delta\rho^z_{\mu\nu}(\omega)$. The latter quantity can be subsequently
transformed to $\delta\rho^z_{ia}(\omega)$ via Eq.~\eqref{eq:rhonn} keeping
only the electron-hole part. Thus, in practical implementation, the linearity
of the equations allows exchanging the order of Fourier transformation and
matrix multiplications.

Finally, we note that in our experience it is advantageous to store the whole
time-dependent evolution of the system, \ie, $C^z_{\mu n}(t)$, as done in the
present implementation. While alternative on-the-fly Fourier or other
transformations would reduce the amount of required storage space, they would
restrict the analysis to the set of parameters specified at the outset of the
calculation.

\subsection{Linear response of the density matrix in the Casida method}
\label{sec:casida}

In Casida's linear-response formulation of TDDFT \cite{Casida1995,Casida2009}
the response is obtained by solving the matrix eigenvalue equation
\begin{equation}
    \M \Omega \M F_I = \omega_I^2 \M F_I
\end{equation}
yielding excitation energies $\omega_I$ and corresponding Casida eigenvectors
$\M F_I$. The matrix $\M \Omega$ is constructed in the KS electron-hole space.
Using a double-index $ia$ ($jb$) to denote a KS excitation from an occupied
state $i$ ($j$) to an unoccupied state $a$ ($b$), the elements of the matrix
can be written as
\begin{equation}
    \Omega_{ia,jb} = \omega_{ia}^2 \delta_{ia, jb} +
    2 \sqrt{f_{ia} \omega_{ia}} K_{ia,jb} \sqrt{f_{jb} \omega_{jb}} ,
\end{equation}
where $f_{ia} = f_a - f_i$ is the occupation number difference,
$\omega_{ia} = \epsilon_a - \epsilon_i$ is the KS eigenvalue difference, see
Eq.~\eqref{eq:gs}, and the matrix $K_{ia,jb}$ represents the coupling between
the excitations $i \to a$ and $j \to b$. \cite{Casida1995}

The linear response of the density matrix at frequency $\omega$ can be obtained
as \cite{Casida1995}
\begin{align}
    \delta\rho^z_{ia}(\omega) =
    \sum_{jb}^{\eh}
    \sqrt{f_{ia} \omega_{ia}} \,
    \left(\M \Omega - \omega^2 \M 1\right)^{-1}_{ia,jb} \,
    \sqrt{f_{jb} \omega_{jb}} \mu^z_{jb} ,
    \label{eq:rho cas inv}
\end{align}
where the summation runs over electron-hole pairs (eh) and involves the dipole
matrix elements
$\mu^z_{jb} = - \int \psi^{(0)*}_b(\V r) z \psi^{(0)}_j(\V r) \, \dee \V r$.
Using the spectral decomposition \cite{Casida1995}
$\left(\M \Omega - \omega^2 \M 1\right)^{-1}_{ia,jb} = \sum_{I} F_{I,ia} G_I(\omega) F_{I,jb}^*$,
where $G_I(\omega) = 1 / (\omega_I^2 - \omega^2)$, allows us to write
Eq.~\eqref{eq:rho cas inv} as
\begin{align}
    \delta\rho^z_{ia}(\omega) =
    \sqrt{f_{ia} \omega_{ia}}
    \sum_{I} F_{I,ia} G_I(\omega)
    \sum_{jb}^{\eh}
    F_{I,jb}^*
    \sqrt{f_{jb} \omega_{jb}} \mu^z_{jb} .
    \label{eq:rho cas}
\end{align}
The term $G_I(\omega)$ is divergent at excitation energies $\omega_I$ in the
common TDDFT implementations due to the infinite lifetime of the excitations.
Analogously to the time domain, a finite lifetime for the excitations can be
restored by the substitution $\omega \to \omega + i \eta$, where the arbitrary
parameter $\eta$ determines the lifetime. This leads to a Lorentzian line
shape and the imaginary part is given by
\begin{align}
    \text{Im}\left[G_I(\omega)\right] = \frac{\pi}{2\omega_I} \left[L(\omega) - L(-\omega)\right],
    \label{eq:imag part}
\end{align}
where $L(\omega) = 1/\pi \cdot \eta/[(\omega - \omega_I)^2 + \eta^2]$ is the
Lorentzian function. Alternatively, the Gaussian line shape can be obtained by
using the Gaussian function
$g(\omega)=1/\sqrt{2\pi}\sigma \cdot \exp[-(\omega - \omega_I)^2/2\sigma^2]$
instead of the Lorentzian function $L(\omega)$ in Eq.~\eqref{eq:imag part}.

\subsection{Kohn--Sham decomposition}

The linear response of the density matrix in the KS electron-hole space,
$\delta\rho^z_{ia}(\omega)$, can be calculated equivalently using both the
real-time propagation [Eq.~\eqref{eq:rho td}] and the Casida approach
[Eq.~\eqref{eq:rho cas}]. While this quantity would already allow the analysis
of the response at frequency $\omega$ in terms of its components in the KS
electron-hole space, a more intuitive analysis can be obtained by connecting
$\delta\rho^z_{ia}(\omega)$ to an observable photo-absorption cross-section
describing the resonances of the system. First, the dynamical polarizability
is given by\cite{Casida1995}
\begin{align}
    \alpha_{xz}(\omega) &= 2 \sum_{ia}^{\eh} \mu^{x*}_{ia} \delta\rho^z_{ia}(\omega) .
    \label{eq:pol ia}
\end{align}
and the photo-absorption is described the dipole strength function
\begin{align}
    S_{z}(\omega) = \frac{2 \omega}{\pi} \Im\left[\alpha_{zz}(\omega)\right] ,
    \label{eq:abs}
\end{align}
which is normalized to integrate to the number of electrons in the system
$N_e$, \ie, $\int_{0}^{\infty} S_{z}(\omega) \dee \omega = N_{e}$.
This is similar to the sum rule $\sum_I f^z_I = N_e$, where
$f^z_I = 2 \left| \sum_{ia} \mu^{z*}_{ia} \sqrt{f_{ia} \omega_{ia}} F_{I,ia}\right|^2$
is the oscillator strength of the discrete excitation $I$. \cite{Casida1995}

By comparing Eqs.~\eqref{eq:pol ia} and \eqref{eq:abs},
we can now define the KS decomposition of the absorption spectrum as
\begin{align}
    S^{z}_{ia}(\omega) = \frac{4 \omega}{\pi} \Im\left[\mu^{z*}_{ia} \delta\rho^z_{ia}(\omega)\right] .
    \label{eq:abs ia}
\end{align}
This quantity is used to analyze the response of silver nanoparticles in
Sec.~\ref{sec:ag} below. Previously, similar photo-absorption decompositions
have been used in the electron-hole space \cite{Repisky2015} and based on, \eg,
spatial location \cite{Koval2016, Kolesov2016} or angular momentum. \cite{Koval2016}

\section{Results}
\label{sec:results}

\subsection{Benzene derivatives}
\label{sec:ben}

To benchmark the presented methods and their computational implementation, we
now analyze the optical response of the molecular systems benzene (\ce{C6H6}),
naphthalene (\ce{C10H8}), and anthracene (\ce{C14H10}) using both the RT-TDDFT
and Casida implementations in \textlcsc{GPAW} package.
\cite{Mortensen2005, Enkovaara2010, GPAW, Walter2008}
These characteristic conjugated molecules are suited for the present benchmark as
they have well-defined $\pi \rightarrow \pi^*$ transitions that exhibit a
systematic red-shift as the extent of the conjugated $\pi$-system increases.
\cite{Wilkinson1956, Ferguson1957}

As the real-time propagation uses the full time-dependent Hamiltonian matrices,
the end result includes contributions from all electron-hole pairs and the
limit of the full KS space is automatically achieved by propagating only the
occupied orbitals. This is in contrast to the \textlcsc{GPAW} implementation
of the Casida approach, \cite{Walter2008} which commonly requires setting an
energy cut-off that determines the KS transitions included in the calculation
of the Casida matrix. In order to ensure comparability of the results, we have
included in the calculation of the Casida matrix all the transitions that are
possible within the KS electron-hole space spanned by the orbitals.

Both the RT-TDDFT and Casida calculations were carried out using the default
PAW data sets and the default double-$\zeta$ polarized (dzp) basis sets within
the LCAO description. While these dzp basis sets might not be sufficient for
yielding numerical values at the complete-basis-set limit,
\cite{Larsen2009, Rossi2015Nanoplasmonics}
they are suitable for qualitative analyses and for the benchmarking study
presented here. The Perdew-Burke-Ernzerhof (PBE) \cite{Perdew1996,*Perdew1997}
exchange-correlation functional was employed in the adiabatic limit. A coarse
grid spacing of $0.3$~\AA\ was chosen to represent densities and potentials and
the molecules were surrounded by a vacuum region of at least 6\,\AA. The
Hartree potential was evaluated with a multigrid Poisson solver using the
monopole and dipole corrections for the potential.

For the RT-TDDFT calculations, we used a small time step of
$\Delta t = 5\,\text{as}$ in order to achieve high numerical accuracy. The
total propagation time was $T=30\,\text{fs}$, which is sufficient for the used
Gaussian broadening with $\sigma = 0.07\,\text{eV}$ corresponding to a full
width at half-maximum (FWHM) of $0.16\,\text{eV}$.

The calculated photo-absorption spectra of the benzene derivatives are shown in
Fig.~\ref{fig:ben-spec}. The Casida and RT-TDDFT methods yield virtually
indistinguishable spectra. For conciseness, we only present an analysis for
excitations along the long axis ($x$) of the molecules. Note, however, that
the response in the other directions can be analyzed in similar fashion.

\subsubsection*{Casida approach}

\begin{figure}[t]
    \includegraphics[scale=1]{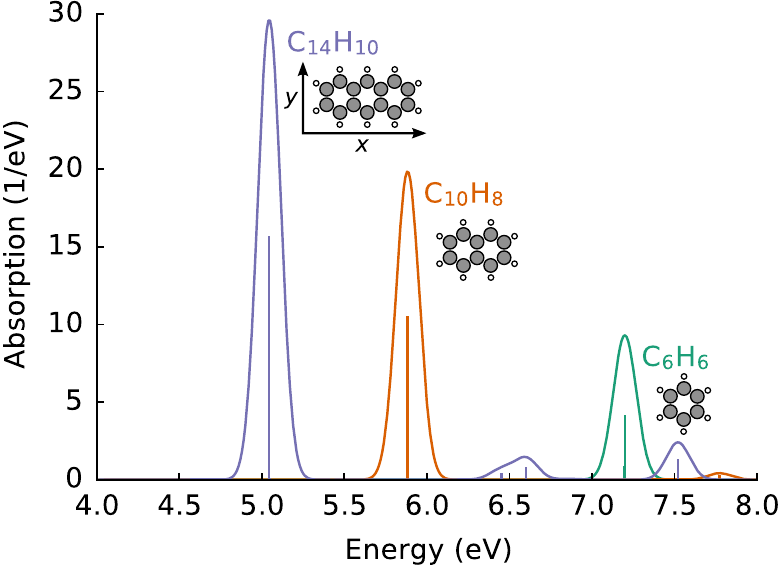}
    \caption{Photo-absorption spectra $S_{x}(\omega)$
    along the long axis ($x$) of the benzene derivatives.}
    \label{fig:ben-spec}
\end{figure}

The response of each of the molecules is dominated by a single absorption peak
(see Fig.~\ref{fig:ben-spec}), which results from discrete excitations. In
Table~\ref{tab:ben-cas}, we show the KS decomposition of these excitations as
described by the components of the normalized Casida eigenvectors $F_{I,ia}$.
Due to the normalization, $\sum_{ia} F_{I,ia}^2 = 1$ for each excitation $I$.

For benzene (\ce{C6H6}, point group $\text{D}_{6\text{h}}$) the excitation at
7.2\,eV corresponds to the first $\text{E}_\text{1u}$ transition from the
doubly degenerate highest occupied molecular orbital
(HOMO; $\text{E}_\text{1g}$) to the doubly degenerate lowest unoccupied
molecular orbital (LUMO; $\text{E}_\text{2u}$). In the present calculations
the symmetry of the molecule has not been enforced and the orbitals
$\pi_{-0/1}$ and $\pi^*_{+0/1}$ span the $\text{E}_\text{1g}$ and
$\text{E}_\text{2u}$ symmetries, respectively. Implementation-dependent
numerical factors slightly lift their degeneracy and determine the exact
unitary rotation between the states.

\begin{table}[t]
    \caption{%
    Casida analysis of the most prominent excitations of benzene (\ce{C6H6}),
    naphthalene (\ce{C10H8}), and naphthalene (\ce{C14H10}). Orbitals are
    enumerated with respect to HOMO ($\pi_{-0}$) and LUMO ($\pi^*_{+0}$). The
    orbital characters are given in brackets based on the point groups
    $\text{D}_\text{6h}$ (benzene) and $\text{D}_\text{2h}$ (naphthalene,
    anthracene).
    }
    \label{tab:ben-cas}
    \begin{ruledtabular}
\begin{tabular}{lllD{.}{\,\to\,}{-1}l}
Molecule & $\omega_I$ (eV) & $f_I^x$ & i . a & $F^2_{I,ia}$ \\
\hline
\multirow{8}{*}{      C$_{6}$H$_{6}$} & \multirow{4}{*}{$   7.198$} & \multirow{4}{*}{$    0.2784$} &   
  \pi_{-1}(\text{E}_\text{1g}) .   \pi^*_{+1}(\text{E}_\text{2u}) &  0.31430 \\
                                      &                             &                               &   
  \pi_{-0}(\text{E}_\text{1g}) .   \pi^*_{+0}(\text{E}_\text{2u}) &  0.31254 \\
                                      &                             &                               &   
  \pi_{-1}(\text{E}_\text{1g}) .   \pi^*_{+0}(\text{E}_\text{2u}) &  0.16863 \\
                                      &                             &                               &   
  \pi_{-0}(\text{E}_\text{1g}) .   \pi^*_{+1}(\text{E}_\text{2u}) &  0.16833 \\
\cline{2-5}
                                      & \multirow{4}{*}{$   7.199$} & \multirow{4}{*}{$    1.3546$} &   
  \pi_{-1}(\text{E}_\text{1g}) .   \pi^*_{+0}(\text{E}_\text{2u}) &  0.31362 \\
                                      &                             &                               &   
  \pi_{-0}(\text{E}_\text{1g}) .   \pi^*_{+1}(\text{E}_\text{2u}) &  0.31325 \\
                                      &                             &                               &   
  \pi_{-1}(\text{E}_\text{1g}) .   \pi^*_{+1}(\text{E}_\text{2u}) &  0.16895 \\
                                      &                             &                               &   
  \pi_{-0}(\text{E}_\text{1g}) .   \pi^*_{+0}(\text{E}_\text{2u}) &  0.16793 \\
\hline
\multirow{2}{*}{     C$_{10}$H$_{8}$} & \multirow{2}{*}{$   5.883$} & \multirow{2}{*}{$    3.4839$} &   
  \pi_{-0}(\text{A}_\text{u})  .   \pi^*_{+1}(\text{B}_\text{3g}) &  0.48451 \\
                                      &                             &                               &   
  \pi_{-1}(\text{B}_\text{2u}) .   \pi^*_{+0}(\text{B}_\text{1g}) &  0.47748 \\
\hline
\multirow{3}{*}{    C$_{14}$H$_{10}$} & \multirow{3}{*}{$   5.044$} & \multirow{3}{*}{$    5.2000$} &   
  \pi_{-0}(\text{B}_\text{3g}) .   \pi^*_{+1}(\text{A}_\text{u})  &  0.50237 \\
                                      &                             &                               &   
  \pi_{-1}(\text{B}_\text{2g}) .   \pi^*_{+0}(\text{B}_\text{1u}) &  0.45773 \\
                                      &                             &                               &   
  \pi_{-4}(\text{B}_\text{1u}) .   \pi^*_{+2}(\text{B}_\text{2g}) &  0.01049 \\
\end{tabular}
    \end{ruledtabular}
\end{table}

Naphthalene (\ce{C10H8}) and anthracene (\ce{C14H10}) belong to the
$\text{D}_{2\text{h}}$ symmetry point group. In both molecules the most
prominent excitation is the $\text{B}_\text{3u}$ transition, which is mainly
composed of transitions from HOMO to LUMO$+1$ and HOMO$-1$ to LUMO. While in
naphthalene the other contributions amount to less than 1\%, in anthracene, a
minor contribution originates also from a transition from HOMO$-4$ to LUMO$+2$.

\subsubsection*{RT-TDDFT approach}

The Casida eigenvector $F_{I,ia}$ considered in Table~\ref{tab:ben-cas} is
directly related to the linear response of the density matrix, see
Eq.~\eqref{eq:rho cas}, and is employed here for benchmarking the RT-TDDFT
methodology described Sec.~\ref{sec:rttdft}. In order to proceed with
comparison, consider a discrete excitation $J$ that is energetically separated
from other excitations. Since $\Im[G_I(\omega_J)]$ in Eq.~\eqref{eq:imag part}
is approximately zero when $I \neq J$, only the excitation $J$ contributes in
Eq.~\eqref{eq:rho cas}. This implies that
$\Im[\rho^x_{ia}(\omega_J)] \approx A \sqrt{f_{ia} \omega_{ia}} F_{J,ia}$,
where $A$ is a constant independent of index $ia$. Thus, after normalization,
$\Im[\rho^x_{ia}(\omega_J)] / \sqrt{f_{ia} \omega_{ia}} \equiv F^x_{ia}(\omega_J)$
yields the components of the Casida eigenvector $F_{J,ia}$. This connection
allows us to calculate the Casida eigenvector also from the RT-TDDFT approach.
This is demonstrated in Table~\ref{tab:ben-rt}, in which we show the calculated
KS decompositions at the peak energies of the photo-absorption spectrum
(Fig.~\ref{fig:ben-spec}).

\begin{table}[t]
    \caption{%
    RT-TDDFT analysis at the peak energies $\omega$ of benzene (\ce{C6H6}),
    naphthalene (\ce{C10H8}), and naphthalene (\ce{C14H10}). The intensities
    $S_x(\omega)$ have been multiplied with the area under the peak to facilitate a
    comparison with the oscillator strengths $f^x_I$ shown in
    Table~\ref{tab:ben-cas}. The last column shows for reference
    $[F^{x}_{ia}(\omega)]^2$ as calculated with the Casida approach.
    }
    \label{tab:ben-rt}
    \begin{ruledtabular}
\begin{tabular}{lllD{.}{\,\to\,}{-1}ll}
Molecule & $\omega$ (eV) & $S_{x}(\omega)$ & i . a & $[F^{x}_{ia}(\omega)]^2$ & Casida  \\
\hline
\multirow{4}{*}{      C$_{6}$H$_{6}$} & \multirow{4}{*}{$    7.20$} & \multirow{4}{*}{$  1.6283$} &   
  \pi_{-1} .   \pi^*_{+0} &  0.46184 &  0.46186 \\
                                      &                             &                             &   
  \pi_{-0} .   \pi^*_{+1} &  0.46126 &  0.46126 \\
                                      &                             &                             &   
  \pi_{-1} .   \pi^*_{+1} &  0.02045 &  0.02043 \\
                                      &                             &                             &   
  \pi_{-0} .   \pi^*_{+0} &  0.02032 &  0.02030 \\
\hline
\multirow{2}{*}{     C$_{10}$H$_{8}$} & \multirow{2}{*}{$    5.88$} & \multirow{2}{*}{$  3.4776$} &   
  \pi_{-0} .   \pi^*_{+1} &  0.48472 &  0.48451 \\
                                      &                             &                             &   
  \pi_{-1} .   \pi^*_{+0} &  0.47728 &  0.47748 \\
\hline
\multirow{3}{*}{    C$_{14}$H$_{10}$} & \multirow{3}{*}{$    5.04$} & \multirow{3}{*}{$  5.1903$} &   
  \pi_{-0} .   \pi^*_{+1} &  0.50277 &  0.50241 \\
                                      &                             &                             &   
  \pi_{-1} .   \pi^*_{+0} &  0.45745 &  0.45777 \\
                                      &                             &                             &   
  \pi_{-4} .   \pi^*_{+2} &  0.01044 &  0.01049 \\
\end{tabular}
    \end{ruledtabular}
\end{table}

In the case of benzene (\ce{C6H6}), we inevitably obtain a superposition of the
two underlying degenerate excitations (see Table~\ref{tab:ben-cas}). We can,
however, calculate the equivalent superimposed $F^x_{ia}(\omega)$ eigenvector
also from the Casida approach (shown in the last column of
Table~\ref{tab:ben-rt}). For this quantity, we obtain an excellent match
between the RT-TDDFT and Casida approaches.

For naphthalene (\ce{C10H8}) and anthracene (\ce{C14H10}), a single excitation
dominates the response and $F_{I,ia}^2$ and $[F^x_{ia}(\omega)]^2$ should yield
the same decomposition as discussed above. Indeed, we observe that the
RT-TDDFT calculations of the decomposition $[F^x_{ia}(\omega)]^2$ reproduce the
discrete Casida eigenvector $F_{I,ia}^2$ with very good numerical accuracy.
When both $F_{I,ia}^2$ and $[F^x_{ia}(\omega)]^2$ are calculated with the
Casida approach, their values should be identical if the excitation is
completely isolated. While for naphthalene, these quantities are exactly the
same up to the shown number of digits (compare the last columns of
Tables~\ref{tab:ben-cas} and \ref{tab:ben-rt}), for anthracene, the numerical
values differ slightly. This deviation is due to a small contribution from a
weak excitation that is close in energy
($\omega_I = 5.051\,\text{eV}$, $f^x_I = 5 \cdot 10^{-4}$)
to the dominant excitation of the anthracene molecule.

\subsection{Silver nanoparticles}
\label{sec:ag}

TDDFT calculations of noble metal nanoparticles up to diameters of several
nanometers are computationally demanding, but the have become feasible with
recent developments. \cite{Iida2014, Kuisma2015, Baseggio2015, Baseggio2016,
Koval2016} Here, we focus on silver nanoparticles as prototypical nanoplasmonic
systems with a strong plasmonic response in the visible--ultraviolet light
regime. \cite{Scholl2012, Haberland2013} Using the methodology described above
in conjunction with a recent RT-TDDFT implementation \cite{Kuisma2015}, we can
analyze the response of silver nanoparticles with reasonable computational
resources. For illustration, a full real-time propagation of 3000~time steps
for \ce{Ag561} can be realized in 110~hours using 144~cores on an Intel Haswell
based architecture. \cite{NoteTaito}

Kuisma \etal\ have previously studied icosahedral silver nanoparticles composed
of 55, 147, 309, and 561 atoms corresponding to diameters ranging from $1.1$~nm
to $2.7$~nm \cite{Kuisma2015}. Here, we consider the same nanoparticle series
and use the same geometries and computational parameters as in
Ref.~\onlinecite{Kuisma2015}. We employ optimized LCAO basis sets
\cite{Kuisma2015} and the orbital-dependent Gritsenko-van Leeuwen-van
Lenthe-Baerends (GLLB) \cite{Gritsenko1995} exchange-correlation potential with
the solid-state modification by Kuisma \etal\ (GLLB-SC) \cite{Kuisma2010},
which yields an accurate description of the $d$ electron states in noble
metals. \cite{Yan2011First, Yan2012, Kuisma2015}

\begin{figure}[t]
    \includegraphics[scale=1]{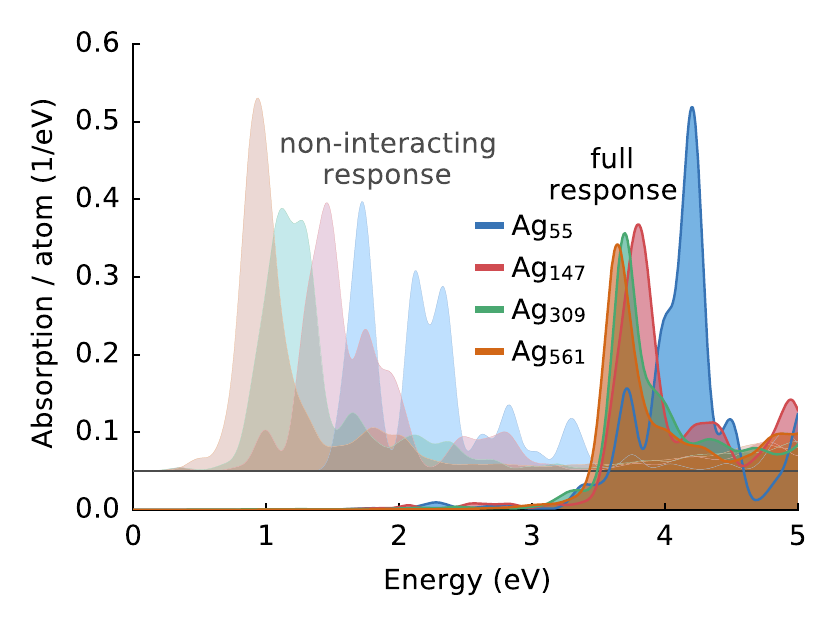}
    \caption{%
    Photo-absorption spectra of icosahedral silver nanoparticles.
    The non-interacting-electron spectra shown for comparison
    are vertically shifted and scaled by a factor of $0.2$.
    }
    \label{fig:ag-spec}
\end{figure}

The calculated photo-absorption spectra of the nanoparticles are shown in
Fig.~\ref{fig:ag-spec}. The non-interacting-electron spectra calculated from
the KS eigenvalue differences $\omega_{ia}$ and transition dipole matrix
elements $\mu^x_{ia}$ are also shown to facilitate the discussion below. In
Ref.~\onlinecite{Kuisma2015} it was observed that the plasmon resonance is
well-formed in \ce{Ag147} and in larger nanoparticles, whereas the response of
\ce{Ag55} consists of multiple peaks, the origin of which cannot be readily
resolved. In the following, we analyze the response of nanoparticles in terms
of the KS decomposition, which enables us to shed light on the response of the
\ce{Ag55} nanoparticle.

\subsubsection*{Transition contribution maps}

\begin{figure*}[t!]
    \subfloat{%
        \includegraphics[scale=1]{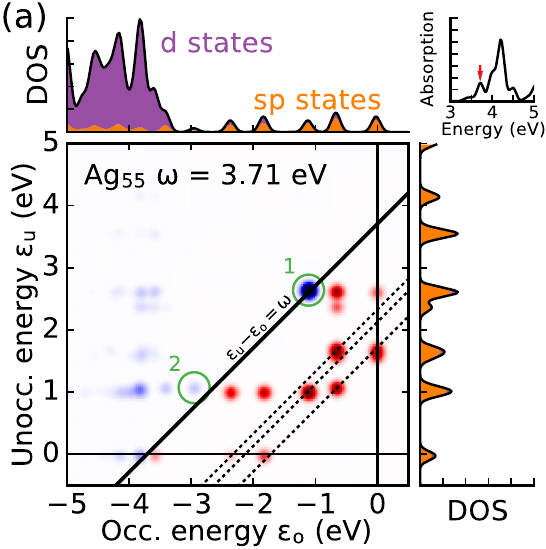}%
        \hfill%
        \includegraphics[scale=1]{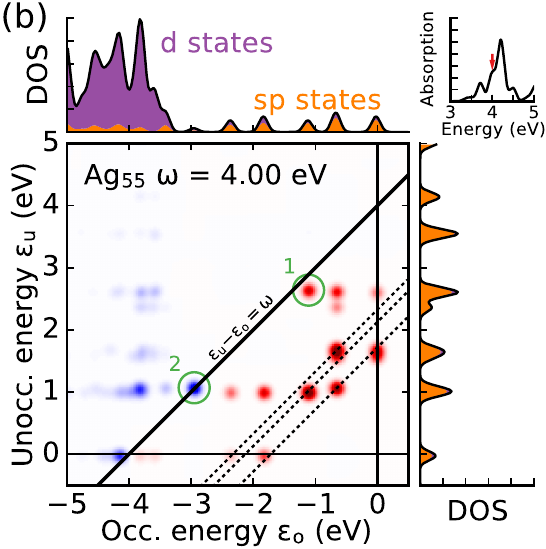}%
        \hfill%
        \includegraphics[scale=1]{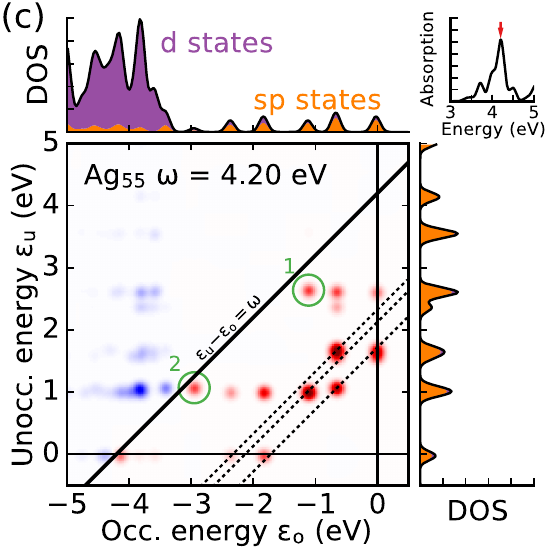}%
    }%
    \\%
    \subfloat{%
        \includegraphics[scale=1]{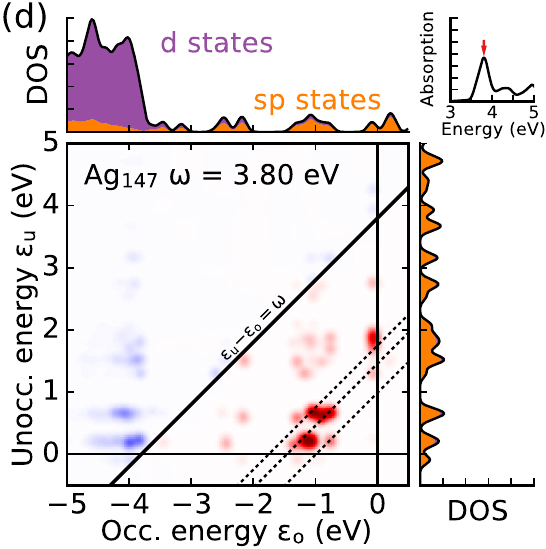}%
        \hfill%
        \includegraphics[scale=1]{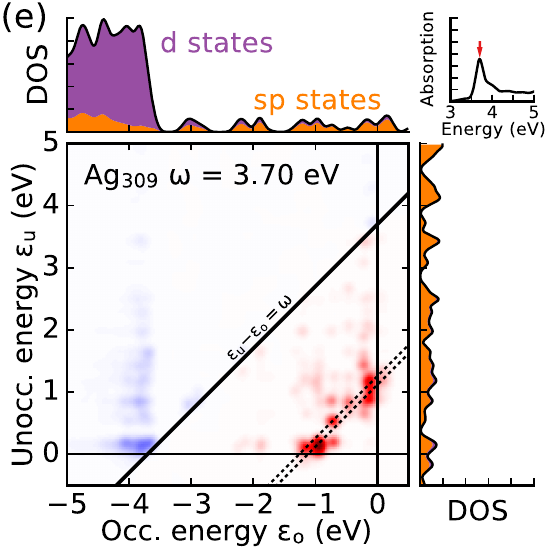}%
        \hfill%
        \includegraphics[scale=1]{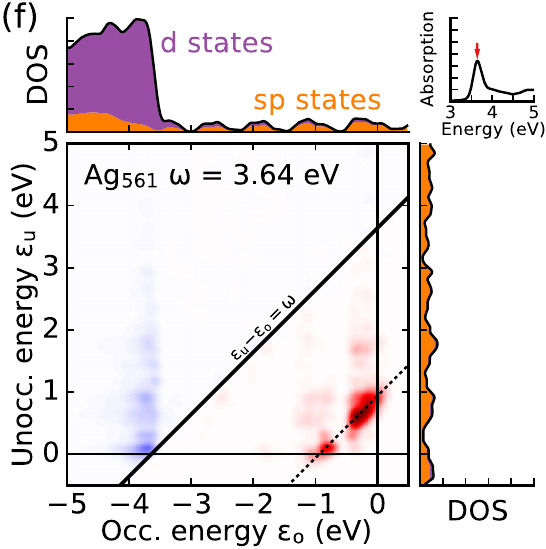}%
    }%
    \caption{%
    Transition contribution maps for the photo-absorption decomposition of
    \ce{Ag55} at different resonance energies $\omega$ (a--c), and those of
    \ce{Ag147} (d), \ce{Ag309} (e), and \ce{Ag561} (f) at the respective plasmon
    resonance energies. The KS eigenvalues are given with respect to the Fermi
    level. The constant transition energy lines
    $\varepsilon_{\text{u}} - \varepsilon_{\text{o}} = \omega$
    are superimposed at the analysis energy (solid line) and at the resonance
    energies of the non-interacting-electron spectra (dashed lines, see
    Fig.~\ref{fig:ag-spec}). Red and blue colors indicate positive and negative
    values of the photo-absorption decomposition, respectively. The inset of each
    panel shows the absorption spectrum with the arrow pointing at the analysis
    frequency $\omega$. The densities of states (DOS) have been colored to indicate
    $sp$ and $d$ character of the states. The transitions marked with green circles
    in panels (a--c) are discussed in the text.
    }
    \label{fig:tcm}
\end{figure*}

In order to analyze the response in terms of the Kohn--Sham decomposition, we
present the decomposition as a transition contribution map (TCM; see
Fig.~\ref{fig:tcm} below) \cite{Malola2013,He2010}, which is an especially
useful representation for plasmonic systems in which resonances are typically
superpositions of many electron-hole excitations. The TCM represents the KS
decomposition weight $w_{ia}(\omega)$ at a fixed $\omega$ in the
two-dimensional (2D) plane spanned by the energy axes for occupied and
unoccupied states. More specifically, the 2D plot is defined by
\begin{equation}
    M^\text{TCM}_\omega(\varepsilon_{\text{o}}, \varepsilon_{\text{u}}) =
    \sum_{ia} w_{ia}(\omega) g_{ia}(\varepsilon_{\text{o}}, \varepsilon_{\text{u}}),
\end{equation}
where $g_{ia}$ is a 2D broadening function of the discrete KS states.
Here, we employ the Gaussian function
\begin{equation}
    g_{ia}(\varepsilon_{\text{o}}, \varepsilon_{\text{u}}) = ({\sqrt{2 \pi} \sigma})^{-2}
    e^{{-\frac{(\varepsilon_{\text{o}} - \epsilon_i)^2 + (\varepsilon_{\text{u}}- \epsilon_a)^2}{2\sigma^2}}}
\end{equation}
with $\sigma = 0.07\,\text{eV}$. The same $\sigma$ parameter is also used in
the spectral broadening. For the weight $w_{ia}(\omega)$, we use the
absorption decomposition of Eq.~\eqref{eq:abs ia} normalized by the total
absorption, \ie,
\begin{equation}
    w_{ia}(\omega) = S^x_{ia}(\omega)/S_x(\omega) .
    \label{eq:tcm w}
\end{equation}
Due to the icosahedral symmetry of the nanoparticles their response is
isotropic, $S_x(\omega) = S_y(\omega) = S_z(\omega)$, and the decomposition is
degenerate (compare the case of benzene in Sec.~\ref{sec:ben}).

Alternatively, instead of Eq.~\eqref{eq:tcm w} one could use, \eg, the
normalized transition density matrix
($w_{ia}(\omega) = |\rho^x_{ia}(\omega)|^2$)
as the weight. Equation~\eqref{eq:tcm w}, however, has the advantage that it
retains the information about the sign of the response in the KS decomposition
and has a physically sound interpretation as the photo-absorption
decomposition.

TCMs of the nanoparticles at different resonance energies are shown in
Fig.~\ref{fig:tcm} along with the density of states (DOS), which has been
colored to indicate the $sp$ and $d$ character of the states. The latter
decomposition is based on the angular momentum quantum number $l_\mu$ of the
LCAO basis functions indexed by $\mu$. For example, the $d$ character of the
$n$th state is estimated as $\sum_{\mu: l_\mu = 2} | C^{(0)}_{\mu n}|^2$, where
the coefficients are normalized such that
$\sum_{\mu} | C^{(0)}_{\mu n}|^2 = 1$.

\subsubsection*{Analysis of Ag$_{147}$, Ag$_{309}$, and Ag$_{561}$}

First, we consider the largest nanoparticles \ce{Ag147}, \ce{Ag309}, and
\ce{Ag561}, the TCMs of which are shown in Figs.~\ref{fig:tcm}(d--f). The
TCMs highlight two major features in their response. First, there is a strong
positive constructive contribution \cite{Guidez2012} (red features in
Fig.~\ref{fig:tcm}) from the KS transitions whose eigenvalue differences are
significantly lower than the plasmon resonance energy $\omega$. The same
low-energy $sp$ transitions are responsible for the strong peaks in the
non-interacting-electron spectra (see Fig.~\ref{fig:ag-spec}), which are
indicated in Fig.~\ref{fig:tcm} by dashed lines. Thus, TCM shows how the
resonance energy is blue-shifted as the interaction is turned on from the
non-interacting case ($\lambda = 0$) to the fully interacting one
($\lambda = 1$). This demonstrates the plasmonic nature of the excitation in
the so-called $\lambda$-scaling approach for plasmon identification,
\cite{Bernadotte2013, Krauter2015} and illustrates the importance of low-energy
transitions for plasmon formation. \cite{Ma2015} Another prominent feature in
the response is the damping due to $d$ electrons, which is seen in the TCMs as
large negative contributions from occupied $d$ states into unoccupied states
(blue features at $\varepsilon_\text{o}\approx-4\,\text{eV}$ in
Fig.~\ref{fig:tcm}). Interestingly, the plasmon peak appears close to the
onset of $d$ electron transitions, corresponding to the intersection of the
line $\varepsilon_\text{u} - \varepsilon_\text{o} = \omega$ and the horizontal
Fermi level line. Generally, with increasing nanoparticle size the DOS becomes
increasingly continuous, which is also visible in the increasing uniformity of
the TCMs.

In Ref.~\onlinecite{Baseggio2016}, TCMs for charged silver nanoparticles up to
\ce{Ag309} have been studied. The two main features in Fig.~\ref{fig:tcm}, the
low-energy $sp$ transitions and the $d$ electron damping, are in agreement with
these TCMs reported earlier. In contrast to Fig.~\ref{fig:tcm}, the TCMs in
Ref.~\onlinecite{Baseggio2016} show, however, also a significant contribution
from $sp$ transitions close to the
$\varepsilon_\text{u} - \varepsilon_\text{o} = \omega$ line. We consider this
to be due to the different choice of the TCM weight $w_{ia}(\omega)$ in
Ref.~\onlinecite{Baseggio2016}. In the absorption decomposition we used in
Fig.~\ref{fig:tcm} [Eqs.~\eqref{eq:abs ia} and \eqref{eq:tcm w}] the KS
components are essentially weighted with the dipole matrix element
$\mu^x_{ia}$, which affects the relative magnitudes observed in TCM.

\subsubsection*{Analysis of Ag$_{55}$}

Next, we consider the \ce{Ag55} nanoparticle that exhibits multiple strong
peaks in the absorption spectrum, resulting in difficulties in identifying the
plasmon resonance. The TCM analyses for the three prominent peak energies are
shown in Figs.~\ref{fig:tcm}(a--c). Due to its small size, \ce{Ag55} has well
defined, discrete KS states as visible in DOS. The overall features in TCMs
are similar to those of the larger nanoparticles, \ie, the low-energy $sp$
transitions and the $d$ electron transitions yield positive and negative
contributions, respectively, though the low-energy transitions that form the
plasmon are energetically clearly separated.

In contrast to the larger nanoparticles, in the \ce{Ag55} nanoparticle some of
the strongly contributing $sp$ transitions are located close to the peak
frequencies, \ie, close to the
$\varepsilon_{\text{u}} - \varepsilon_{\text{o}} = \omega$
lines in the TCMs. These excitations are marked in Figs.~\ref{fig:tcm}(a--c)
by green circles numbered as 1 and 2. By examining these KS transitions as a
function of frequency $\omega$
(TCMs with the 0.01\,eV resolution are provided as Supplemental Material \cite{NoteSupplement}),
we note that the first transition changes its sign at $\omega = 3.85$\,eV,
close to the minimum between the peak maxima at 3.71\,eV
[Fig.~\ref{fig:tcm}(a)] and 4.00\,eV (b). Similarly, the second transition
changes its sign at $\omega = 4.06$\,eV between the maxima at 4.00\,eV (b) and
4.20\,eV (c). At the same time, the low-energy transitions forming the plasmon
remain mainly unchanged over this frequency window. Thus, the presence of
multiple peaks in the \ce{Ag55} spectrum seems to correspond to a strong
coupling between the marked KS transitions and the plasmon. This is seen as the
splitting of the plasmon into multiple resonances with antisymmetric and
symmetric combinations of the KS transition and the plasmonic transitions. In
the larger nanoparticles, the interaction between the plasmon and the nearby KS
transitions is weak and the coupling is merely seen as a broadening of the
plasmon peak.

A detailed inspection reveals that some $d$ electron transitions also change
their sign in the frequency range where the peak splitting occurs. The changes
in their sign, however, do not match the maxima and minima of the absorption
spectrum like in the case of the marked KS transitions. Thus, we expect the
marked $sp$ transitions to be the major cause for the plasmon splitting.

In the literature, \ce{Ag55} has been reported to have slightly varying spectra
depending, \eg, on the exact geometry, the exchange-correlation functional, and
the numerical parameters used.
\cite{Weissker2011, Bae2012, Rabilloud2014, Rossi2015Nanoplasmonics, Ma2015, Baseggio2016, Koval2016}
Correspondingly, the electronic structures are different and the \ce{Ag55}
spectra have single or multiple peaks. We expect, however, that the splitting
behavior observed here can be a useful general concept for understanding the
response of small plasmonic nanoparticles.

\section{Discussion}
\label{sec:discussion}

The RT-TDDFT approach provides a more favorable scaling with the system size
than the Casida approach. The latter, however, achieves a smaller pre-factor,
especially when using non-local (e.g., hybrid exchange-correlation functionals)
\cite{Sander2017}, which renders it computationally more efficient for small
and moderately-sized systems. In contrast, the RT-TDDFT approach becomes very
attractive for systems comprising thousands of electrons (and typically
hundreds of atoms) such as the silver nanoparticles considered in the present
work. Previously, the lack of a decomposition scheme on par with the Casida
method has been identified as a drawback of the RT-TDDFT approach.
\cite{Sander2017} Here, we have introduced and demonstrated the performance of
a method that overcomes this limitation and represents an efficient tool for
analyzing electronic excitations within RT-TDDFT in general, and plasmonic
response in particular.

It should be noted that in the RT-TDDFT approach the observable response is
sensitive to the external perturbation used to initialize the time propagation.
If the perturbation is chosen to be, say, a dipole perturbation along the $x$
direction, only excitations with a dipole component parallel to $x$ are
observable in the response. By combining at most three separate
time-propagation calculations (possibly even less in the cases of higher
symmetry) with dipole perturbations along the $x$, $y$, and $z$ axes, one can
recover the full dynamical polarizability tensor. However, for obtaining
optically dark (dipole-forbidden) excitations from RT-TDDFT calculations, one
would need to run the time propagation with different initial perturbations.
This is in contrast to the Casida approach, where also dipole-forbidden
excitations are obtained by diagonalizing the $\M \Omega$ matrix.

It was illustrated in Sec.~\ref{sec:ben} that the RT-TDDFT method does not
yield direct access to the discrete spectrum, but rather allows an analysis at
chosen frequencies yielding the combined response coming from all the
contributing discrete excitations. Usually, this is not a significant
restriction as in experimental measurements the energy resolution is limited by
instrumental broadening and the excitation lifetimes. Computationally, the
energy resolution is determined by the broadening parameter, which can be
always reduced by increasing the propagation time. Furthermore, for larger
systems that are the primary application area for RT-TDDFT, the electronic
spectrum becomes increasingly dense and the distinction of individual
excitations is less relevant.

\section{Conclusions}
\label{sec:conclusions}

In this work we have presented that the linear response of the density matrix
in the Kohn--Sham electron-hole basis can be obtained from real-time
propagation TDDFT via a basis transformation. The methodology has been
implemented in a recent RT-TDDFT code \cite{Kuisma2015} and is to be made publicly
available as a part of the open source electronic structure code \textlcsc{GPAW}.
\cite{Mortensen2005, Enkovaara2010, GPAW}

The present approach provides access to the same information via RT-TDDFT that
is usually available only with the Casida approach. This was specifically
demonstrated by a careful comparison of the results for benzene derivatives,
which were shown to be numerically almost identical for the Casida and RT-TDDFT
calculations.

Using the presented methodology, we analyzed the plasmonic response of icosahedral
silver nanoparticles in the Kohn--Sham electron-hole space. The \ce{Ag55}
nanoparticle was considered in detail and the multiple resonances in its
response were shown to reflect the splitting of the plasmon due to the strong
coupling between the plasmon and individual single-electron transitions. In
the larger \ce{Ag147}, \ce{Ag309}, and \ce{Ag561} nanoparticles, the
interaction between plasmon and individual single-electron transitions close to
the resonance is weaker and a distinct plasmon resonance emerges from the
constructive superposition of the low-energy Kohn--Sham transitions
\cite{Bernadotte2013, Krauter2015, Ma2015} accompanied by the damping due to
$d$ electron transitions.

In summary, the present work raises the analysis capabilities of the RT-TDDFT
to the same level as with the Casida approach, without compromising the
computational benefits of RT-TDDFT.

\begin{acknowledgments}
We thank the Academy of Finland for support through its
Centres of Excellence Programme (2012--2017) under
Projects No.~251748 and No.~284621.
M.\! K.\ is grateful for Academy of Finland Postdoctoral Researcher funding
under Project No.~295602.
T.\! P.\! R.\ thanks
the Vilho, Yrj\"{o} and Kalle V\"{a}is\"{a}l\"{a} Foundation of
the Finnish Academy of Science and Letters, and
Finnish Cultural Foundation
for support.
We also thank the Swedish Research Council, the Knut and Alice Wallenberg
Foundation, and the Swedish Foundation for Strategic Research for support.
We acknowledge computational resources provided by
CSC -- IT Center for Science (Finland),
the Aalto Science-IT project (Aalto University School of Science),
the Swedish National Infrastructure for Computing at NSC (Link\"oping) and at PDC (Stockholm).
\end{acknowledgments}

\appendix*

\section{Derivation of Eq.~\eqref{eq:rhonn} within the PAW formalism}
\label{app:derivation}

Within the PAW formalism, Eq.~\eqref{eq:tdks} reads
\begin{equation}
    i \mathcal{T}^\dagger \mathcal{T} \frac{\partial}{\partial t} \widetilde \psi_n(\V r,t) =
    \mathcal{T}^\dagger H_{\text{KS}}(t) \mathcal{T} \widetilde \psi_n(\V r,t) ,
\end{equation}
where $\widetilde \psi_n(\V r,t)$ is a pseudo wave function and
$\mathcal{T}$ denotes the PAW transformation operator \cite{Blochl1994}.

In the LCAO method, the pseudo wave function $\widetilde \psi_n(\V r, t)$ is
expanded in localized basis functions $\widetilde \phi_\mu(\V r)$ centered at
atomic coordinates
\begin{equation}
    \widetilde \psi_n(\V r, t) = \sum_\mu \widetilde \phi_\mu(\V r) C_{\mu n}(t),
\end{equation}
with expansion coefficients $C_{\mu n}(t)$.
The corresponding all-electron wave function is given by [compare to
Eq.~\eqref{eq:lcaowf}]
\begin{equation}
    \psi_n(\V r, t) = \mathcal{T} \widetilde \psi_n(\V r, t)
     = \sum_\mu \phi_\mu(\V r) C_{\mu n}(t),
    \label{eq:wf}
\end{equation}
where the all-electron basis functions have been defined as
$\phi_\mu = \mathcal{T} \widetilde \phi_\mu$.

The time-dependent all-electron real-space density matrix can be obtained as
\begin{equation}
    \rho(\V r, \V r', t) = \sum_{\mu\nu} \phi_\mu(\V r) \rho_{\mu\nu}(t) \phi_\nu^*(\V r') ,
    \label{eq:denmat}
\end{equation}
where the density matrix in the LCAO basis $\rho_{\mu\nu}(t)$ is given by
Eq.~\eqref{eq:rhomm}.

The transformation of the real-space density matrix to the basis defined by the
ground-state KS orbitals $\psi_{n}^{(0)}(\V r)$, see Eq.~\eqref{eq:gs}, is
given by
\begin{equation}
    \rho_{nn'}(t) = \int {\rm d} \V r \int {\rm d} \V r' \psi_{n}^{(0)*}(\V r) \rho(\V r, \V r',t) \psi_{n'}^{(0)}(\V r').
    \label{eq:rhonn1}
\end{equation}
By expanding $\psi_n^{(0)}(\V r)$ in the LCAO basis as in Eq.~\eqref{eq:wf}
and inserting Eq.~\eqref{eq:denmat} into Eq.~\eqref{eq:rhonn1}, we obtain
after reordering the integrals
\begin{align}
    \rho_{nn'}(t) &= \sum_{\mu} C^{(0)*}_{\mu n} \sum_{\mu'} \underbrace{\int \dee \V r \phi_{\mu}^*(\V r) \phi_{\mu'}(\V r)}_{S_{\mu \mu'}} \nonumber \\
    & \cdot
    \sum_{\nu'} \rho_{\mu'\nu'}(t)
    \sum_\nu \underbrace{\int \dee \V r' \phi_{\nu'}^*(\V r') \phi_{\nu}(\V r')}_{S^*_{\nu \nu'}} C^{(0)}_{\nu n'}
    .
    \label{eq:rhonn2}
\end{align}
Here, we have isolated the overlap integrals $S_{\mu\mu'}$ used regularly in
LCAO calculations, \ie,
\begin{equation}
    S_{\mu\mu'} =
    \int \dee \V r \phi^*_{\mu}(\V r) \phi_{\mu'}(\V r) =
    \int \dee \V r \widetilde\phi^*_{\mu}(\V r) \mathcal{T}^\dagger \mathcal{T} \widetilde\phi_{\mu'}(\V r).
    \label{eq:overlap}
\end{equation}
After simplifying the overlap integrals in Eq.~\eqref{eq:rhonn2}, we obtain
Eq.~\eqref{eq:rhonn}. We note that the PAW transformation affects only the
evaluation of the overlap integrals $S_{\mu\mu'}$, see Eq.~\eqref{eq:overlap}.

\bibliography{article}

\begin{thebibliography}{112}%
\makeatletter
\providecommand \@ifxundefined [1]{%
 \@ifx{#1\undefined}
}%
\providecommand \@ifnum [1]{%
 \ifnum #1\expandafter \@firstoftwo
 \else \expandafter \@secondoftwo
 \fi
}%
\providecommand \@ifx [1]{%
 \ifx #1\expandafter \@firstoftwo
 \else \expandafter \@secondoftwo
 \fi
}%
\providecommand \natexlab [1]{#1}%
\providecommand \enquote  [1]{``#1''}%
\providecommand \bibnamefont  [1]{#1}%
\providecommand \bibfnamefont [1]{#1}%
\providecommand \citenamefont [1]{#1}%
\providecommand \href@noop [0]{\@secondoftwo}%
\providecommand \href [0]{\begingroup \@sanitize@url \@href}%
\providecommand \@href[1]{\@@startlink{#1}\@@href}%
\providecommand \@@href[1]{\endgroup#1\@@endlink}%
\providecommand \@sanitize@url [0]{\catcode `\\12\catcode `\$12\catcode
  `\&12\catcode `\#12\catcode `\^12\catcode `\_12\catcode `\%12\relax}%
\providecommand \@@startlink[1]{}%
\providecommand \@@endlink[0]{}%
\providecommand \url  [0]{\begingroup\@sanitize@url \@url }%
\providecommand \@url [1]{\endgroup\@href {#1}{\urlprefix }}%
\providecommand \urlprefix  [0]{URL }%
\providecommand \Eprint [0]{\href }%
\providecommand \doibase [0]{http://dx.doi.org/}%
\providecommand \selectlanguage [0]{\@gobble}%
\providecommand \bibinfo  [0]{\@secondoftwo}%
\providecommand \bibfield  [0]{\@secondoftwo}%
\providecommand \translation [1]{[#1]}%
\providecommand \BibitemOpen [0]{}%
\providecommand \bibitemStop [0]{}%
\providecommand \bibitemNoStop [0]{.\EOS\space}%
\providecommand \EOS [0]{\spacefactor3000\relax}%
\providecommand \BibitemShut  [1]{\csname bibitem#1\endcsname}%
\let\auto@bib@innerbib\@empty
\bibitem [{\citenamefont {Runge}\ and\ \citenamefont
  {Gross}(1984)}]{Runge1984}%
  \BibitemOpen
  \bibfield  {author} {\bibinfo {author} {\bibfnamefont {E.}~\bibnamefont
  {Runge}}\ and\ \bibinfo {author} {\bibfnamefont {E.~K.~U.}\ \bibnamefont
  {Gross}},\ }\href {\doibase 10.1103/PhysRevLett.52.997} {\bibfield  {journal}
  {\bibinfo  {journal} {Phys. Rev. Lett.}\ }\textbf {\bibinfo {volume} {52}},\
  \bibinfo {pages} {997} (\bibinfo {year} {1984})}\BibitemShut {NoStop}%
\bibitem [{\citenamefont {Hohenberg}\ and\ \citenamefont
  {Kohn}(1964)}]{Hohenberg1964}%
  \BibitemOpen
  \bibfield  {author} {\bibinfo {author} {\bibfnamefont {P.}~\bibnamefont
  {Hohenberg}}\ and\ \bibinfo {author} {\bibfnamefont {W.}~\bibnamefont
  {Kohn}},\ }\href {\doibase 10.1103/PhysRev.136.B864} {\bibfield  {journal}
  {\bibinfo  {journal} {Phys. Rev.}\ }\textbf {\bibinfo {volume} {136}},\
  \bibinfo {pages} {B864} (\bibinfo {year} {1964})}\BibitemShut {NoStop}%
\bibitem [{\citenamefont {Kohn}\ and\ \citenamefont {Sham}(1965)}]{Kohn1965}%
  \BibitemOpen
  \bibfield  {author} {\bibinfo {author} {\bibfnamefont {W.}~\bibnamefont
  {Kohn}}\ and\ \bibinfo {author} {\bibfnamefont {L.~J.}\ \bibnamefont
  {Sham}},\ }\href {\doibase 10.1103/PhysRev.140.A1133} {\bibfield  {journal}
  {\bibinfo  {journal} {Phys. Rev.}\ }\textbf {\bibinfo {volume} {140}},\
  \bibinfo {pages} {A1133} (\bibinfo {year} {1965})}\BibitemShut {NoStop}%
\bibitem [{\citenamefont {Marques}\ \emph {et~al.}(2012)\citenamefont
  {Marques}, \citenamefont {Maitra}, \citenamefont {Nogueira}, \citenamefont
  {Gross},\ and\ \citenamefont {Rubio}}]{Marques2012}%
  \BibitemOpen
  \bibinfo {editor} {\bibfnamefont {M.~A.~L.}\ \bibnamefont {Marques}},
  \bibinfo {editor} {\bibfnamefont {N.~T.}\ \bibnamefont {Maitra}}, \bibinfo
  {editor} {\bibfnamefont {F.~M.~S.}\ \bibnamefont {Nogueira}}, \bibinfo
  {editor} {\bibfnamefont {E.~K.~U.}\ \bibnamefont {Gross}}, \ and\ \bibinfo
  {editor} {\bibfnamefont {A.}~\bibnamefont {Rubio}},\ eds.,\ \href {\doibase
  10.1007/978-3-642-23518-4} {\emph {\bibinfo {title} {{Fundamentals of
  Time-Dependent Density Functional Theory}}}},\ \bibinfo {series} {Lecture
  Notes in Physics}, Vol.\ \bibinfo {volume} {837}\ (\bibinfo  {publisher}
  {Springer},\ \bibinfo {year} {2012})\BibitemShut {NoStop}%
\bibitem [{\citenamefont {Ullrich}(2012)}]{Ullrich2012}%
  \BibitemOpen
  \bibfield  {author} {\bibinfo {author} {\bibfnamefont {C.~A.}\ \bibnamefont
  {Ullrich}},\ }\href {\doibase 10.1093/acprof:oso/9780199563029.001.0001}
  {\emph {\bibinfo {title} {{Time-Dependent Density-Functional Theory: Concepts
  and Applications}}}}\ (\bibinfo  {publisher} {Oxford University Press},\
  \bibinfo {year} {2012})\BibitemShut {NoStop}%
\bibitem [{\citenamefont {Ekardt}(1984)}]{Ekardt1984}%
  \BibitemOpen
  \bibfield  {author} {\bibinfo {author} {\bibfnamefont {W.}~\bibnamefont
  {Ekardt}},\ }\href {\doibase 10.1103/PhysRevLett.52.1925} {\bibfield
  {journal} {\bibinfo  {journal} {Phys. Rev. Lett.}\ }\textbf {\bibinfo
  {volume} {52}},\ \bibinfo {pages} {1925} (\bibinfo {year}
  {1984})}\BibitemShut {NoStop}%
\bibitem [{\citenamefont {Puska}\ \emph {et~al.}(1985)\citenamefont {Puska},
  \citenamefont {Nieminen},\ and\ \citenamefont {Manninen}}]{Puska1985}%
  \BibitemOpen
  \bibfield  {author} {\bibinfo {author} {\bibfnamefont {M.~J.}\ \bibnamefont
  {Puska}}, \bibinfo {author} {\bibfnamefont {R.~M.}\ \bibnamefont {Nieminen}},
  \ and\ \bibinfo {author} {\bibfnamefont {M.}~\bibnamefont {Manninen}},\
  }\href {\doibase 10.1103/PhysRevB.31.3486} {\bibfield  {journal} {\bibinfo
  {journal} {Phys. Rev. B}\ }\textbf {\bibinfo {volume} {31}},\ \bibinfo
  {pages} {3486} (\bibinfo {year} {1985})}\BibitemShut {NoStop}%
\bibitem [{\citenamefont {Beck}(1987)}]{Beck1987}%
  \BibitemOpen
  \bibfield  {author} {\bibinfo {author} {\bibfnamefont {D.~E.}\ \bibnamefont
  {Beck}},\ }\href {\doibase 10.1103/PhysRevB.35.7325} {\bibfield  {journal}
  {\bibinfo  {journal} {Phys. Rev. B}\ }\textbf {\bibinfo {volume} {35}},\
  \bibinfo {pages} {7325} (\bibinfo {year} {1987})}\BibitemShut {NoStop}%
\bibitem [{\citenamefont {Morton}\ \emph {et~al.}(2011)\citenamefont {Morton},
  \citenamefont {Silverstein},\ and\ \citenamefont {Jensen}}]{Morton2011}%
  \BibitemOpen
  \bibfield  {author} {\bibinfo {author} {\bibfnamefont {S.~M.}\ \bibnamefont
  {Morton}}, \bibinfo {author} {\bibfnamefont {D.~W.}\ \bibnamefont
  {Silverstein}}, \ and\ \bibinfo {author} {\bibfnamefont {L.}~\bibnamefont
  {Jensen}},\ }\href {\doibase 10.1021/cr100265f} {\bibfield  {journal}
  {\bibinfo  {journal} {Chem. Rev.}\ }\textbf {\bibinfo {volume} {111}},\
  \bibinfo {pages} {3962} (\bibinfo {year} {2011})}\BibitemShut {NoStop}%
\bibitem [{\citenamefont {Varas}\ \emph {et~al.}(2016)\citenamefont {Varas},
  \citenamefont {Garc{\'i}a-Gonz{\'a}lez}, \citenamefont {Feist}, \citenamefont
  {Garc{\'i}a-Vidal},\ and\ \citenamefont {Rubio}}]{Varas2016}%
  \BibitemOpen
  \bibfield  {author} {\bibinfo {author} {\bibfnamefont {A.}~\bibnamefont
  {Varas}}, \bibinfo {author} {\bibfnamefont {P.}~\bibnamefont
  {Garc{\'i}a-Gonz{\'a}lez}}, \bibinfo {author} {\bibfnamefont
  {J.}~\bibnamefont {Feist}}, \bibinfo {author} {\bibfnamefont
  {F.}~\bibnamefont {Garc{\'i}a-Vidal}}, \ and\ \bibinfo {author}
  {\bibfnamefont {A.}~\bibnamefont {Rubio}},\ }\href {\doibase
  10.1515/nanoph-2015-0141} {\bibfield  {journal} {\bibinfo  {journal}
  {Nanophotonics}\ }\textbf {\bibinfo {volume} {5}},\ \bibinfo {pages} {409}
  (\bibinfo {year} {2016})}\BibitemShut {NoStop}%
\bibitem [{\citenamefont {Prodan}\ \emph {et~al.}(2003)\citenamefont {Prodan},
  \citenamefont {Nordlander},\ and\ \citenamefont
  {Halas}}]{Prodan2003Electronic}%
  \BibitemOpen
  \bibfield  {author} {\bibinfo {author} {\bibfnamefont {E.}~\bibnamefont
  {Prodan}}, \bibinfo {author} {\bibfnamefont {P.}~\bibnamefont {Nordlander}},
  \ and\ \bibinfo {author} {\bibfnamefont {N.~J.}\ \bibnamefont {Halas}},\
  }\href {\doibase 10.1021/nl034594q} {\bibfield  {journal} {\bibinfo
  {journal} {Nano Lett.}\ }\textbf {\bibinfo {volume} {3}},\ \bibinfo {pages}
  {1411} (\bibinfo {year} {2003})}\BibitemShut {NoStop}%
\bibitem [{\citenamefont {Aikens}\ \emph {et~al.}(2008)\citenamefont {Aikens},
  \citenamefont {Li},\ and\ \citenamefont {Schatz}}]{Aikens2008}%
  \BibitemOpen
  \bibfield  {author} {\bibinfo {author} {\bibfnamefont {C.~M.}\ \bibnamefont
  {Aikens}}, \bibinfo {author} {\bibfnamefont {S.}~\bibnamefont {Li}}, \ and\
  \bibinfo {author} {\bibfnamefont {G.~C.}\ \bibnamefont {Schatz}},\ }\href
  {\doibase 10.1021/jp802707r} {\bibfield  {journal} {\bibinfo  {journal} {J.
  Phys. Chem. C}\ }\textbf {\bibinfo {volume} {112}},\ \bibinfo {pages} {11272}
  (\bibinfo {year} {2008})}\BibitemShut {NoStop}%
\bibitem [{\citenamefont {Zuloaga}\ \emph {et~al.}(2010)\citenamefont
  {Zuloaga}, \citenamefont {Prodan},\ and\ \citenamefont
  {Nordlander}}]{Zuloaga2010}%
  \BibitemOpen
  \bibfield  {author} {\bibinfo {author} {\bibfnamefont {J.}~\bibnamefont
  {Zuloaga}}, \bibinfo {author} {\bibfnamefont {E.}~\bibnamefont {Prodan}}, \
  and\ \bibinfo {author} {\bibfnamefont {P.}~\bibnamefont {Nordlander}},\
  }\href {\doibase 10.1021/nn101589n} {\bibfield  {journal} {\bibinfo
  {journal} {{ACS} Nano}\ }\textbf {\bibinfo {volume} {4}},\ \bibinfo {pages}
  {5269} (\bibinfo {year} {2010})}\BibitemShut {NoStop}%
\bibitem [{\citenamefont {Weissker}\ and\ \citenamefont
  {Mottet}(2011)}]{Weissker2011}%
  \BibitemOpen
  \bibfield  {author} {\bibinfo {author} {\bibfnamefont {H.-C.}\ \bibnamefont
  {Weissker}}\ and\ \bibinfo {author} {\bibfnamefont {C.}~\bibnamefont
  {Mottet}},\ }\href {\doibase 10.1103/PhysRevB.84.165443} {\bibfield
  {journal} {\bibinfo  {journal} {Phys. Rev. B}\ }\textbf {\bibinfo {volume}
  {84}},\ \bibinfo {pages} {165443} (\bibinfo {year} {2011})}\BibitemShut
  {NoStop}%
\bibitem [{\citenamefont {Li}\ \emph {et~al.}(2013)\citenamefont {Li},
  \citenamefont {Hayashi},\ and\ \citenamefont {Guo}}]{Li2013}%
  \BibitemOpen
  \bibfield  {author} {\bibinfo {author} {\bibfnamefont {J.-H.}\ \bibnamefont
  {Li}}, \bibinfo {author} {\bibfnamefont {M.}~\bibnamefont {Hayashi}}, \ and\
  \bibinfo {author} {\bibfnamefont {G.-Y.}\ \bibnamefont {Guo}},\ }\href
  {\doibase 10.1103/PhysRevB.88.155437} {\bibfield  {journal} {\bibinfo
  {journal} {Phys. Rev. B}\ }\textbf {\bibinfo {volume} {88}},\ \bibinfo
  {pages} {155437} (\bibinfo {year} {2013})}\BibitemShut {NoStop}%
\bibitem [{\citenamefont {Piccini}\ \emph {et~al.}(2013)\citenamefont
  {Piccini}, \citenamefont {Havenith}, \citenamefont {Broer},\ and\
  \citenamefont {Stener}}]{Piccini2013}%
  \BibitemOpen
  \bibfield  {author} {\bibinfo {author} {\bibfnamefont {G.}~\bibnamefont
  {Piccini}}, \bibinfo {author} {\bibfnamefont {R.~W.~A.}\ \bibnamefont
  {Havenith}}, \bibinfo {author} {\bibfnamefont {R.}~\bibnamefont {Broer}}, \
  and\ \bibinfo {author} {\bibfnamefont {M.}~\bibnamefont {Stener}},\ }\href
  {\doibase 10.1021/jp405769e} {\bibfield  {journal} {\bibinfo  {journal} {J.
  Phys. Chem. C}\ }\textbf {\bibinfo {volume} {117}},\ \bibinfo {pages} {17196}
  (\bibinfo {year} {2013})}\BibitemShut {NoStop}%
\bibitem [{\citenamefont {Burgess}\ and\ \citenamefont
  {Keast}(2014)}]{Burgess2014}%
  \BibitemOpen
  \bibfield  {author} {\bibinfo {author} {\bibfnamefont {R.~W.}\ \bibnamefont
  {Burgess}}\ and\ \bibinfo {author} {\bibfnamefont {V.~J.}\ \bibnamefont
  {Keast}},\ }\href {\doibase 10.1021/jp408545c} {\bibfield  {journal}
  {\bibinfo  {journal} {J. Phys. Chem. C}\ }\textbf {\bibinfo {volume} {118}},\
  \bibinfo {pages} {3194} (\bibinfo {year} {2014})}\BibitemShut {NoStop}%
\bibitem [{\citenamefont {Barcaro}\ \emph {et~al.}(2014)\citenamefont
  {Barcaro}, \citenamefont {Sementa}, \citenamefont {Fortunelli},\ and\
  \citenamefont {Stener}}]{Barcaro2014}%
  \BibitemOpen
  \bibfield  {author} {\bibinfo {author} {\bibfnamefont {G.}~\bibnamefont
  {Barcaro}}, \bibinfo {author} {\bibfnamefont {L.}~\bibnamefont {Sementa}},
  \bibinfo {author} {\bibfnamefont {A.}~\bibnamefont {Fortunelli}}, \ and\
  \bibinfo {author} {\bibfnamefont {M.}~\bibnamefont {Stener}},\ }\href
  {\doibase 10.1021/jp5016565} {\bibfield  {journal} {\bibinfo  {journal} {J.
  Phys. Chem. C}\ }\textbf {\bibinfo {volume} {118}},\ \bibinfo {pages} {12450}
  (\bibinfo {year} {2014})}\BibitemShut {NoStop}%
\bibitem [{\citenamefont {Weissker}\ and\ \citenamefont
  {Lopez-Lozano}(2015)}]{Weissker2015}%
  \BibitemOpen
  \bibfield  {author} {\bibinfo {author} {\bibfnamefont {H.-C.}\ \bibnamefont
  {Weissker}}\ and\ \bibinfo {author} {\bibfnamefont {X.}~\bibnamefont
  {Lopez-Lozano}},\ }\href {\doibase 10.1039/C5CP01177A} {\bibfield  {journal}
  {\bibinfo  {journal} {Phys. Chem. Chem. Phys.}\ }\textbf {\bibinfo {volume}
  {17}},\ \bibinfo {pages} {28379} (\bibinfo {year} {2015})}\BibitemShut
  {NoStop}%
\bibitem [{\citenamefont {Bae}\ and\ \citenamefont {Aikens}(2015)}]{Bae2015}%
  \BibitemOpen
  \bibfield  {author} {\bibinfo {author} {\bibfnamefont {G.-T.}\ \bibnamefont
  {Bae}}\ and\ \bibinfo {author} {\bibfnamefont {C.~M.}\ \bibnamefont
  {Aikens}},\ }\href {\doibase 10.1021/acs.jpcc.5b05978} {\bibfield  {journal}
  {\bibinfo  {journal} {J. Phys. Chem. C}\ }\textbf {\bibinfo {volume} {119}},\
  \bibinfo {pages} {23127} (\bibinfo {year} {2015})}\BibitemShut {NoStop}%
\bibitem [{\citenamefont {Zapata~Herrera}\ \emph {et~al.}(2016)\citenamefont
  {Zapata~Herrera}, \citenamefont {Aizpurua}, \citenamefont {Kazansky},\ and\
  \citenamefont {Borisov}}]{Zapata2016}%
  \BibitemOpen
  \bibfield  {author} {\bibinfo {author} {\bibfnamefont {M.}~\bibnamefont
  {Zapata~Herrera}}, \bibinfo {author} {\bibfnamefont {J.}~\bibnamefont
  {Aizpurua}}, \bibinfo {author} {\bibfnamefont {A.~K.}\ \bibnamefont
  {Kazansky}}, \ and\ \bibinfo {author} {\bibfnamefont {A.~G.}\ \bibnamefont
  {Borisov}},\ }\href {\doibase 10.1021/acs.langmuir.6b00112} {\bibfield
  {journal} {\bibinfo  {journal} {Langmuir}\ }\textbf {\bibinfo {volume}
  {32}},\ \bibinfo {pages} {2829} (\bibinfo {year} {2016})}\BibitemShut
  {NoStop}%
\bibitem [{\citenamefont {Zuloaga}\ \emph {et~al.}(2009)\citenamefont
  {Zuloaga}, \citenamefont {Prodan},\ and\ \citenamefont
  {Nordlander}}]{Zuloaga2009}%
  \BibitemOpen
  \bibfield  {author} {\bibinfo {author} {\bibfnamefont {J.}~\bibnamefont
  {Zuloaga}}, \bibinfo {author} {\bibfnamefont {E.}~\bibnamefont {Prodan}}, \
  and\ \bibinfo {author} {\bibfnamefont {P.}~\bibnamefont {Nordlander}},\
  }\href {\doibase 10.1021/nl803811g} {\bibfield  {journal} {\bibinfo
  {journal} {Nano Lett.}\ }\textbf {\bibinfo {volume} {9}},\ \bibinfo {pages}
  {887} (\bibinfo {year} {2009})}\BibitemShut {NoStop}%
\bibitem [{\citenamefont {Song}\ \emph {et~al.}(2011)\citenamefont {Song},
  \citenamefont {Nordlander},\ and\ \citenamefont {Gao}}]{Song2011}%
  \BibitemOpen
  \bibfield  {author} {\bibinfo {author} {\bibfnamefont {P.}~\bibnamefont
  {Song}}, \bibinfo {author} {\bibfnamefont {P.}~\bibnamefont {Nordlander}}, \
  and\ \bibinfo {author} {\bibfnamefont {S.}~\bibnamefont {Gao}},\ }\href
  {\doibase 10.1063/1.3554420} {\bibfield  {journal} {\bibinfo  {journal} {J.
  Chem. Phys.}\ }\textbf {\bibinfo {volume} {134}},\ \bibinfo {pages} {074701}
  (\bibinfo {year} {2011})}\BibitemShut {NoStop}%
\bibitem [{\citenamefont {Song}\ \emph {et~al.}(2012)\citenamefont {Song},
  \citenamefont {Meng}, \citenamefont {Nordlander},\ and\ \citenamefont
  {Gao}}]{Song2012}%
  \BibitemOpen
  \bibfield  {author} {\bibinfo {author} {\bibfnamefont {P.}~\bibnamefont
  {Song}}, \bibinfo {author} {\bibfnamefont {S.}~\bibnamefont {Meng}}, \bibinfo
  {author} {\bibfnamefont {P.}~\bibnamefont {Nordlander}}, \ and\ \bibinfo
  {author} {\bibfnamefont {S.}~\bibnamefont {Gao}},\ }\href {\doibase
  10.1103/PhysRevB.86.121410} {\bibfield  {journal} {\bibinfo  {journal} {Phys.
  Rev. B}\ }\textbf {\bibinfo {volume} {86}},\ \bibinfo {pages} {121410}
  (\bibinfo {year} {2012})}\BibitemShut {NoStop}%
\bibitem [{\citenamefont {Marinica}\ \emph {et~al.}(2012)\citenamefont
  {Marinica}, \citenamefont {Kazansky}, \citenamefont {Nordlander},
  \citenamefont {Aizpurua},\ and\ \citenamefont {Borisov}}]{Marinica2012}%
  \BibitemOpen
  \bibfield  {author} {\bibinfo {author} {\bibfnamefont {D.}~\bibnamefont
  {Marinica}}, \bibinfo {author} {\bibfnamefont {A.}~\bibnamefont {Kazansky}},
  \bibinfo {author} {\bibfnamefont {P.}~\bibnamefont {Nordlander}}, \bibinfo
  {author} {\bibfnamefont {J.}~\bibnamefont {Aizpurua}}, \ and\ \bibinfo
  {author} {\bibfnamefont {A.~G.}\ \bibnamefont {Borisov}},\ }\href {\doibase
  10.1021/nl300269c} {\bibfield  {journal} {\bibinfo  {journal} {Nano Lett.}\
  }\textbf {\bibinfo {volume} {12}},\ \bibinfo {pages} {1333} (\bibinfo {year}
  {2012})}\BibitemShut {NoStop}%
\bibitem [{\citenamefont {Zhang}\ \emph {et~al.}(2014)\citenamefont {Zhang},
  \citenamefont {Feist}, \citenamefont {Rubio}, \citenamefont
  {Garc\'ia-Gonz\'alez},\ and\ \citenamefont {Garc\'ia-Vidal}}]{Zhang2014}%
  \BibitemOpen
  \bibfield  {author} {\bibinfo {author} {\bibfnamefont {P.}~\bibnamefont
  {Zhang}}, \bibinfo {author} {\bibfnamefont {J.}~\bibnamefont {Feist}},
  \bibinfo {author} {\bibfnamefont {A.}~\bibnamefont {Rubio}}, \bibinfo
  {author} {\bibfnamefont {P.}~\bibnamefont {Garc\'ia-Gonz\'alez}}, \ and\
  \bibinfo {author} {\bibfnamefont {F.~J.}\ \bibnamefont {Garc\'ia-Vidal}},\
  }\href {\doibase 10.1103/PhysRevB.90.161407} {\bibfield  {journal} {\bibinfo
  {journal} {Phys. Rev. B}\ }\textbf {\bibinfo {volume} {90}},\ \bibinfo
  {pages} {161407} (\bibinfo {year} {2014})}\BibitemShut {NoStop}%
\bibitem [{\citenamefont {Varas}\ \emph {et~al.}(2015)\citenamefont {Varas},
  \citenamefont {Garc{\'i}a-Gonz{\'a}lez}, \citenamefont {Garc{\'i}a-Vidal},\
  and\ \citenamefont {Rubio}}]{Varas2015}%
  \BibitemOpen
  \bibfield  {author} {\bibinfo {author} {\bibfnamefont {A.}~\bibnamefont
  {Varas}}, \bibinfo {author} {\bibfnamefont {P.}~\bibnamefont
  {Garc{\'i}a-Gonz{\'a}lez}}, \bibinfo {author} {\bibfnamefont {F.~J.}\
  \bibnamefont {Garc{\'i}a-Vidal}}, \ and\ \bibinfo {author} {\bibfnamefont
  {A.}~\bibnamefont {Rubio}},\ }\href {\doibase 10.1021/acs.jpclett.5b00573}
  {\bibfield  {journal} {\bibinfo  {journal} {J. Phys. Chem. Lett.}\ }\textbf
  {\bibinfo {volume} {6}},\ \bibinfo {pages} {1891} (\bibinfo {year}
  {2015})}\BibitemShut {NoStop}%
\bibitem [{\citenamefont {Barbry}\ \emph {et~al.}(2015)\citenamefont {Barbry},
  \citenamefont {Koval}, \citenamefont {Marchesin}, \citenamefont {Esteban},
  \citenamefont {Borisov}, \citenamefont {Aizpurua},\ and\ \citenamefont
  {S{\'{a}}nchez-Portal}}]{Barbry2015}%
  \BibitemOpen
  \bibfield  {author} {\bibinfo {author} {\bibfnamefont {M.}~\bibnamefont
  {Barbry}}, \bibinfo {author} {\bibfnamefont {P.}~\bibnamefont {Koval}},
  \bibinfo {author} {\bibfnamefont {F.}~\bibnamefont {Marchesin}}, \bibinfo
  {author} {\bibfnamefont {R.}~\bibnamefont {Esteban}}, \bibinfo {author}
  {\bibfnamefont {A.~G.}\ \bibnamefont {Borisov}}, \bibinfo {author}
  {\bibfnamefont {J.}~\bibnamefont {Aizpurua}}, \ and\ \bibinfo {author}
  {\bibfnamefont {D.}~\bibnamefont {S{\'{a}}nchez-Portal}},\ }\href {\doibase
  10.1021/acs.nanolett.5b00759} {\bibfield  {journal} {\bibinfo  {journal}
  {Nano Lett.}\ }\textbf {\bibinfo {volume} {15}},\ \bibinfo {pages} {3410}
  (\bibinfo {year} {2015})}\BibitemShut {NoStop}%
\bibitem [{\citenamefont {Kulkarni}\ and\ \citenamefont
  {Manjavacas}(2015)}]{Kulkarni2015}%
  \BibitemOpen
  \bibfield  {author} {\bibinfo {author} {\bibfnamefont {V.}~\bibnamefont
  {Kulkarni}}\ and\ \bibinfo {author} {\bibfnamefont {A.}~\bibnamefont
  {Manjavacas}},\ }\href {\doibase 10.1021/acsphotonics.5b00246} {\bibfield
  {journal} {\bibinfo  {journal} {ACS Photonics}\ }\textbf {\bibinfo {volume}
  {2}},\ \bibinfo {pages} {987} (\bibinfo {year} {2015})}\BibitemShut {NoStop}%
\bibitem [{\citenamefont {Rossi}\ \emph
  {et~al.}(2015{\natexlab{a}})\citenamefont {Rossi}, \citenamefont
  {Zugarramurdi}, \citenamefont {Puska},\ and\ \citenamefont
  {Nieminen}}]{Rossi2015Quantized}%
  \BibitemOpen
  \bibfield  {author} {\bibinfo {author} {\bibfnamefont {T.~P.}\ \bibnamefont
  {Rossi}}, \bibinfo {author} {\bibfnamefont {A.}~\bibnamefont {Zugarramurdi}},
  \bibinfo {author} {\bibfnamefont {M.~J.}\ \bibnamefont {Puska}}, \ and\
  \bibinfo {author} {\bibfnamefont {R.~M.}\ \bibnamefont {Nieminen}},\ }\href
  {\doibase 10.1103/PhysRevLett.115.236804} {\bibfield  {journal} {\bibinfo
  {journal} {Phys. Rev. Lett.}\ }\textbf {\bibinfo {volume} {115}},\ \bibinfo
  {pages} {236804} (\bibinfo {year} {2015}{\natexlab{a}})}\BibitemShut
  {NoStop}%
\bibitem [{\citenamefont {Marchesin}\ \emph {et~al.}(2016)\citenamefont
  {Marchesin}, \citenamefont {Koval}, \citenamefont {Barbry}, \citenamefont
  {Aizpurua},\ and\ \citenamefont {S{\'a}nchez-Portal}}]{Marchesin2016}%
  \BibitemOpen
  \bibfield  {author} {\bibinfo {author} {\bibfnamefont {F.}~\bibnamefont
  {Marchesin}}, \bibinfo {author} {\bibfnamefont {P.}~\bibnamefont {Koval}},
  \bibinfo {author} {\bibfnamefont {M.}~\bibnamefont {Barbry}}, \bibinfo
  {author} {\bibfnamefont {J.}~\bibnamefont {Aizpurua}}, \ and\ \bibinfo
  {author} {\bibfnamefont {D.}~\bibnamefont {S{\'a}nchez-Portal}},\ }\href
  {\doibase 10.1021/acsphotonics.5b00609} {\bibfield  {journal} {\bibinfo
  {journal} {ACS Photonics}\ }\textbf {\bibinfo {volume} {3}},\ \bibinfo
  {pages} {269} (\bibinfo {year} {2016})}\BibitemShut {NoStop}%
\bibitem [{\citenamefont {Lahtinen}\ \emph {et~al.}(2016)\citenamefont
  {Lahtinen}, \citenamefont {Hulkko}, \citenamefont {Sokolowska}, \citenamefont
  {Tero}, \citenamefont {Saarnio}, \citenamefont {Lindgren}, \citenamefont
  {Pettersson}, \citenamefont {Hakkinen},\ and\ \citenamefont
  {Lehtovaara}}]{Lahtinen2016}%
  \BibitemOpen
  \bibfield  {author} {\bibinfo {author} {\bibfnamefont {T.}~\bibnamefont
  {Lahtinen}}, \bibinfo {author} {\bibfnamefont {E.}~\bibnamefont {Hulkko}},
  \bibinfo {author} {\bibfnamefont {K.}~\bibnamefont {Sokolowska}}, \bibinfo
  {author} {\bibfnamefont {T.-R.}\ \bibnamefont {Tero}}, \bibinfo {author}
  {\bibfnamefont {V.}~\bibnamefont {Saarnio}}, \bibinfo {author} {\bibfnamefont
  {J.}~\bibnamefont {Lindgren}}, \bibinfo {author} {\bibfnamefont
  {M.}~\bibnamefont {Pettersson}}, \bibinfo {author} {\bibfnamefont
  {H.}~\bibnamefont {Hakkinen}}, \ and\ \bibinfo {author} {\bibfnamefont
  {L.}~\bibnamefont {Lehtovaara}},\ }\href {\doibase 10.1039/C6NR05267C}
  {\bibfield  {journal} {\bibinfo  {journal} {Nanoscale}\ }\textbf {\bibinfo
  {volume} {8}},\ \bibinfo {pages} {18665} (\bibinfo {year}
  {2016})}\BibitemShut {NoStop}%
\bibitem [{\citenamefont {Manjavacas}\ \emph {et~al.}(2013)\citenamefont
  {Manjavacas}, \citenamefont {Marchesin}, \citenamefont {Thongrattanasiri},
  \citenamefont {Koval}, \citenamefont {Nordlander}, \citenamefont
  {S{\'a}nchez-Portal},\ and\ \citenamefont {Garc{\'i}a~de
  Abajo}}]{Manjavacas2013}%
  \BibitemOpen
  \bibfield  {author} {\bibinfo {author} {\bibfnamefont {A.}~\bibnamefont
  {Manjavacas}}, \bibinfo {author} {\bibfnamefont {F.}~\bibnamefont
  {Marchesin}}, \bibinfo {author} {\bibfnamefont {S.}~\bibnamefont
  {Thongrattanasiri}}, \bibinfo {author} {\bibfnamefont {P.}~\bibnamefont
  {Koval}}, \bibinfo {author} {\bibfnamefont {P.}~\bibnamefont {Nordlander}},
  \bibinfo {author} {\bibfnamefont {D.}~\bibnamefont {S{\'a}nchez-Portal}}, \
  and\ \bibinfo {author} {\bibfnamefont {F.~J.}\ \bibnamefont {Garc{\'i}a~de
  Abajo}},\ }\href {\doibase 10.1021/nn4006297} {\bibfield  {journal} {\bibinfo
   {journal} {ACS Nano}\ }\textbf {\bibinfo {volume} {7}},\ \bibinfo {pages}
  {3635} (\bibinfo {year} {2013})}\BibitemShut {NoStop}%
\bibitem [{\citenamefont {Andersen}\ and\ \citenamefont
  {Thygesen}(2013)}]{Andersen2013Plasmons}%
  \BibitemOpen
  \bibfield  {author} {\bibinfo {author} {\bibfnamefont {K.}~\bibnamefont
  {Andersen}}\ and\ \bibinfo {author} {\bibfnamefont {K.~S.}\ \bibnamefont
  {Thygesen}},\ }\href {\doibase 10.1103/PhysRevB.88.155128} {\bibfield
  {journal} {\bibinfo  {journal} {Phys. Rev. B}\ }\textbf {\bibinfo {volume}
  {88}},\ \bibinfo {pages} {155128} (\bibinfo {year} {2013})}\BibitemShut
  {NoStop}%
\bibitem [{\citenamefont {Andersen}\ \emph {et~al.}(2014)\citenamefont
  {Andersen}, \citenamefont {Jacobsen},\ and\ \citenamefont
  {Thygesen}}]{Andersen2014}%
  \BibitemOpen
  \bibfield  {author} {\bibinfo {author} {\bibfnamefont {K.}~\bibnamefont
  {Andersen}}, \bibinfo {author} {\bibfnamefont {K.~W.}\ \bibnamefont
  {Jacobsen}}, \ and\ \bibinfo {author} {\bibfnamefont {K.~S.}\ \bibnamefont
  {Thygesen}},\ }\href {\doibase 10.1103/PhysRevB.90.161410} {\bibfield
  {journal} {\bibinfo  {journal} {Phys. Rev. B}\ }\textbf {\bibinfo {volume}
  {90}},\ \bibinfo {pages} {161410} (\bibinfo {year} {2014})}\BibitemShut
  {NoStop}%
\bibitem [{\citenamefont {Lauchner}\ \emph {et~al.}(2015)\citenamefont
  {Lauchner}, \citenamefont {Schlather}, \citenamefont {Manjavacas},
  \citenamefont {Cui}, \citenamefont {McClain}, \citenamefont {Stec},
  \citenamefont {Garc{\'i}a~de Abajo}, \citenamefont {Nordlander},\ and\
  \citenamefont {Halas}}]{Lauchner2015}%
  \BibitemOpen
  \bibfield  {author} {\bibinfo {author} {\bibfnamefont {A.}~\bibnamefont
  {Lauchner}}, \bibinfo {author} {\bibfnamefont {A.~E.}\ \bibnamefont
  {Schlather}}, \bibinfo {author} {\bibfnamefont {A.}~\bibnamefont
  {Manjavacas}}, \bibinfo {author} {\bibfnamefont {Y.}~\bibnamefont {Cui}},
  \bibinfo {author} {\bibfnamefont {M.~J.}\ \bibnamefont {McClain}}, \bibinfo
  {author} {\bibfnamefont {G.~J.}\ \bibnamefont {Stec}}, \bibinfo {author}
  {\bibfnamefont {F.~J.}\ \bibnamefont {Garc{\'i}a~de Abajo}}, \bibinfo
  {author} {\bibfnamefont {P.}~\bibnamefont {Nordlander}}, \ and\ \bibinfo
  {author} {\bibfnamefont {N.~J.}\ \bibnamefont {Halas}},\ }\href {\doibase
  10.1021/acs.nanolett.5b02549} {\bibfield  {journal} {\bibinfo  {journal}
  {Nano Lett.}\ }\textbf {\bibinfo {volume} {15}},\ \bibinfo {pages} {6208}
  (\bibinfo {year} {2015})}\BibitemShut {NoStop}%
\bibitem [{\citenamefont {Gao}\ and\ \citenamefont {Yuan}(2005)}]{Gao2005}%
  \BibitemOpen
  \bibfield  {author} {\bibinfo {author} {\bibfnamefont {S.}~\bibnamefont
  {Gao}}\ and\ \bibinfo {author} {\bibfnamefont {Z.}~\bibnamefont {Yuan}},\
  }\href {\doibase 10.1103/PhysRevB.72.121406} {\bibfield  {journal} {\bibinfo
  {journal} {Phys. Rev. B}\ }\textbf {\bibinfo {volume} {72}},\ \bibinfo
  {pages} {121406} (\bibinfo {year} {2005})}\BibitemShut {NoStop}%
\bibitem [{\citenamefont {Yan}\ \emph {et~al.}(2007)\citenamefont {Yan},
  \citenamefont {Yuan},\ and\ \citenamefont {Gao}}]{Yan2007}%
  \BibitemOpen
  \bibfield  {author} {\bibinfo {author} {\bibfnamefont {J.}~\bibnamefont
  {Yan}}, \bibinfo {author} {\bibfnamefont {Z.}~\bibnamefont {Yuan}}, \ and\
  \bibinfo {author} {\bibfnamefont {S.}~\bibnamefont {Gao}},\ }\href {\doibase
  10.1103/PhysRevLett.98.216602} {\bibfield  {journal} {\bibinfo  {journal}
  {Phys. Rev. Lett.}\ }\textbf {\bibinfo {volume} {98}},\ \bibinfo {pages}
  {216602} (\bibinfo {year} {2007})}\BibitemShut {NoStop}%
\bibitem [{\citenamefont {Bernadotte}\ \emph {et~al.}(2013)\citenamefont
  {Bernadotte}, \citenamefont {Evers},\ and\ \citenamefont
  {Jacob}}]{Bernadotte2013}%
  \BibitemOpen
  \bibfield  {author} {\bibinfo {author} {\bibfnamefont {S.}~\bibnamefont
  {Bernadotte}}, \bibinfo {author} {\bibfnamefont {F.}~\bibnamefont {Evers}}, \
  and\ \bibinfo {author} {\bibfnamefont {C.~R.}\ \bibnamefont {Jacob}},\ }\href
  {\doibase 10.1021/jp3113073} {\bibfield  {journal} {\bibinfo  {journal} {J.
  Phys. Chem. C}\ }\textbf {\bibinfo {volume} {117}},\ \bibinfo {pages} {1863}
  (\bibinfo {year} {2013})}\BibitemShut {NoStop}%
\bibitem [{\citenamefont {Malola}\ \emph {et~al.}(2013)\citenamefont {Malola},
  \citenamefont {Lehtovaara}, \citenamefont {Enkovaara},\ and\ \citenamefont
  {H{\"a}kkinen}}]{Malola2013}%
  \BibitemOpen
  \bibfield  {author} {\bibinfo {author} {\bibfnamefont {S.}~\bibnamefont
  {Malola}}, \bibinfo {author} {\bibfnamefont {L.}~\bibnamefont {Lehtovaara}},
  \bibinfo {author} {\bibfnamefont {J.}~\bibnamefont {Enkovaara}}, \ and\
  \bibinfo {author} {\bibfnamefont {H.}~\bibnamefont {H{\"a}kkinen}},\ }\href
  {\doibase 10.1021/nn4046634} {\bibfield  {journal} {\bibinfo  {journal} {ACS
  Nano}\ }\textbf {\bibinfo {volume} {7}},\ \bibinfo {pages} {10263} (\bibinfo
  {year} {2013})}\BibitemShut {NoStop}%
\bibitem [{\citenamefont {Guidez}\ and\ \citenamefont
  {Aikens}(2012)}]{Guidez2012}%
  \BibitemOpen
  \bibfield  {author} {\bibinfo {author} {\bibfnamefont {E.~B.}\ \bibnamefont
  {Guidez}}\ and\ \bibinfo {author} {\bibfnamefont {C.~M.}\ \bibnamefont
  {Aikens}},\ }\href {\doibase 10.1039/C2NR30253E} {\bibfield  {journal}
  {\bibinfo  {journal} {Nanoscale}\ }\textbf {\bibinfo {volume} {4}},\ \bibinfo
  {pages} {4190} (\bibinfo {year} {2012})}\BibitemShut {NoStop}%
\bibitem [{\citenamefont {Guidez}\ and\ \citenamefont
  {Aikens}(2014)}]{Guidez2014}%
  \BibitemOpen
  \bibfield  {author} {\bibinfo {author} {\bibfnamefont {E.~B.}\ \bibnamefont
  {Guidez}}\ and\ \bibinfo {author} {\bibfnamefont {C.~M.}\ \bibnamefont
  {Aikens}},\ }\href {\doibase 10.1039/C4NR02225D} {\bibfield  {journal}
  {\bibinfo  {journal} {Nanoscale}\ }\textbf {\bibinfo {volume} {6}},\ \bibinfo
  {pages} {11512} (\bibinfo {year} {2014})}\BibitemShut {NoStop}%
\bibitem [{\citenamefont {Yasuike}\ \emph {et~al.}(2011)\citenamefont
  {Yasuike}, \citenamefont {Nobusada},\ and\ \citenamefont
  {Hayashi}}]{Yasuike2011}%
  \BibitemOpen
  \bibfield  {author} {\bibinfo {author} {\bibfnamefont {T.}~\bibnamefont
  {Yasuike}}, \bibinfo {author} {\bibfnamefont {K.}~\bibnamefont {Nobusada}}, \
  and\ \bibinfo {author} {\bibfnamefont {M.}~\bibnamefont {Hayashi}},\ }\href
  {\doibase 10.1103/PhysRevA.83.013201} {\bibfield  {journal} {\bibinfo
  {journal} {Phys. Rev. A}\ }\textbf {\bibinfo {volume} {83}},\ \bibinfo
  {pages} {013201} (\bibinfo {year} {2011})}\BibitemShut {NoStop}%
\bibitem [{\citenamefont {Casanova}\ \emph {et~al.}(2016)\citenamefont
  {Casanova}, \citenamefont {Matxain},\ and\ \citenamefont
  {Ugalde}}]{Casanova2016}%
  \BibitemOpen
  \bibfield  {author} {\bibinfo {author} {\bibfnamefont {D.}~\bibnamefont
  {Casanova}}, \bibinfo {author} {\bibfnamefont {J.~M.}\ \bibnamefont
  {Matxain}}, \ and\ \bibinfo {author} {\bibfnamefont {J.~M.}\ \bibnamefont
  {Ugalde}},\ }\href {\doibase 10.1021/acs.jpcc.6b03210} {\bibfield  {journal}
  {\bibinfo  {journal} {J. Phys. Chem. C}\ }\textbf {\bibinfo {volume} {120}},\
  \bibinfo {pages} {12742} (\bibinfo {year} {2016})}\BibitemShut {NoStop}%
\bibitem [{\citenamefont {Townsend}\ and\ \citenamefont
  {Bryant}(2012)}]{Townsend2012}%
  \BibitemOpen
  \bibfield  {author} {\bibinfo {author} {\bibfnamefont {E.}~\bibnamefont
  {Townsend}}\ and\ \bibinfo {author} {\bibfnamefont {G.~W.}\ \bibnamefont
  {Bryant}},\ }\href {\doibase 10.1021/nl2037613} {\bibfield  {journal}
  {\bibinfo  {journal} {Nano Lett.}\ }\textbf {\bibinfo {volume} {12}},\
  \bibinfo {pages} {429} (\bibinfo {year} {2012})}\BibitemShut {NoStop}%
\bibitem [{\citenamefont {Townsend}\ and\ \citenamefont
  {Bryant}(2014)}]{Townsend2014}%
  \BibitemOpen
  \bibfield  {author} {\bibinfo {author} {\bibfnamefont {E.}~\bibnamefont
  {Townsend}}\ and\ \bibinfo {author} {\bibfnamefont {G.~W.}\ \bibnamefont
  {Bryant}},\ }\href {\doibase 10.1088/2040-8978/16/11/114022} {\bibfield
  {journal} {\bibinfo  {journal} {J. Opt.}\ }\textbf {\bibinfo {volume} {16}},\
  \bibinfo {pages} {114022} (\bibinfo {year} {2014})}\BibitemShut {NoStop}%
\bibitem [{\citenamefont {Ma}\ \emph {et~al.}(2015)\citenamefont {Ma},
  \citenamefont {Wang},\ and\ \citenamefont {Wang}}]{Ma2015}%
  \BibitemOpen
  \bibfield  {author} {\bibinfo {author} {\bibfnamefont {J.}~\bibnamefont
  {Ma}}, \bibinfo {author} {\bibfnamefont {Z.}~\bibnamefont {Wang}}, \ and\
  \bibinfo {author} {\bibfnamefont {L.-W.}\ \bibnamefont {Wang}},\ }\href
  {\doibase 10.1038/ncomms10107} {\bibfield  {journal} {\bibinfo  {journal}
  {Nat. Commun.}\ }\textbf {\bibinfo {volume} {6}},\ \bibinfo {pages} {10107}
  (\bibinfo {year} {2015})}\BibitemShut {NoStop}%
\bibitem [{\citenamefont {Bursi}\ \emph {et~al.}(2016)\citenamefont {Bursi},
  \citenamefont {Calzolari}, \citenamefont {Corni},\ and\ \citenamefont
  {Molinari}}]{Bursi2016}%
  \BibitemOpen
  \bibfield  {author} {\bibinfo {author} {\bibfnamefont {L.}~\bibnamefont
  {Bursi}}, \bibinfo {author} {\bibfnamefont {A.}~\bibnamefont {Calzolari}},
  \bibinfo {author} {\bibfnamefont {S.}~\bibnamefont {Corni}}, \ and\ \bibinfo
  {author} {\bibfnamefont {E.}~\bibnamefont {Molinari}},\ }\href {\doibase
  10.1021/acsphotonics.5b00688} {\bibfield  {journal} {\bibinfo  {journal} {ACS
  Photonics}\ }\textbf {\bibinfo {volume} {3}},\ \bibinfo {pages} {520}
  (\bibinfo {year} {2016})}\BibitemShut {NoStop}%
\bibitem [{\citenamefont {Esteban}\ \emph {et~al.}(2012)\citenamefont
  {Esteban}, \citenamefont {Borisov}, \citenamefont {Nordlander},\ and\
  \citenamefont {Aizpurua}}]{Esteban2012}%
  \BibitemOpen
  \bibfield  {author} {\bibinfo {author} {\bibfnamefont {R.}~\bibnamefont
  {Esteban}}, \bibinfo {author} {\bibfnamefont {A.~G.}\ \bibnamefont
  {Borisov}}, \bibinfo {author} {\bibfnamefont {P.}~\bibnamefont {Nordlander}},
  \ and\ \bibinfo {author} {\bibfnamefont {J.}~\bibnamefont {Aizpurua}},\
  }\href {\doibase 10.1038/ncomms1806} {\bibfield  {journal} {\bibinfo
  {journal} {Nat. Commun.}\ }\textbf {\bibinfo {volume} {3}},\ \bibinfo {pages}
  {825} (\bibinfo {year} {2012})}\BibitemShut {NoStop}%
\bibitem [{\citenamefont {Stella}\ \emph {et~al.}(2013)\citenamefont {Stella},
  \citenamefont {Zhang}, \citenamefont {Garc{\'i}a-Vidal}, \citenamefont
  {Rubio},\ and\ \citenamefont {Garc{\'i}a-Gonz{\'a}lez}}]{Stella2013}%
  \BibitemOpen
  \bibfield  {author} {\bibinfo {author} {\bibfnamefont {L.}~\bibnamefont
  {Stella}}, \bibinfo {author} {\bibfnamefont {P.}~\bibnamefont {Zhang}},
  \bibinfo {author} {\bibfnamefont {F.~J.}\ \bibnamefont {Garc{\'i}a-Vidal}},
  \bibinfo {author} {\bibfnamefont {A.}~\bibnamefont {Rubio}}, \ and\ \bibinfo
  {author} {\bibfnamefont {P.}~\bibnamefont {Garc{\'i}a-Gonz{\'a}lez}},\ }\href
  {\doibase 10.1021/jp401887y} {\bibfield  {journal} {\bibinfo  {journal} {J.
  Phys. Chem. C}\ }\textbf {\bibinfo {volume} {117}},\ \bibinfo {pages} {8941}
  (\bibinfo {year} {2013})}\BibitemShut {NoStop}%
\bibitem [{\citenamefont {Chen}\ \emph {et~al.}(2015)\citenamefont {Chen},
  \citenamefont {Moore}, \citenamefont {Zekarias},\ and\ \citenamefont
  {Jensen}}]{Chen2015}%
  \BibitemOpen
  \bibfield  {author} {\bibinfo {author} {\bibfnamefont {X.}~\bibnamefont
  {Chen}}, \bibinfo {author} {\bibfnamefont {J.~E.}\ \bibnamefont {Moore}},
  \bibinfo {author} {\bibfnamefont {M.}~\bibnamefont {Zekarias}}, \ and\
  \bibinfo {author} {\bibfnamefont {L.}~\bibnamefont {Jensen}},\ }\href
  {\doibase 10.1038/ncomms9921} {\bibfield  {journal} {\bibinfo  {journal}
  {Nat. Commun.}\ }\textbf {\bibinfo {volume} {6}},\ \bibinfo {pages} {8921}
  (\bibinfo {year} {2015})}\BibitemShut {NoStop}%
\bibitem [{\citenamefont {Yan}\ \emph {et~al.}(2015)\citenamefont {Yan},
  \citenamefont {Wubs},\ and\ \citenamefont {Asger~Mortensen}}]{Yan2015}%
  \BibitemOpen
  \bibfield  {author} {\bibinfo {author} {\bibfnamefont {W.}~\bibnamefont
  {Yan}}, \bibinfo {author} {\bibfnamefont {M.}~\bibnamefont {Wubs}}, \ and\
  \bibinfo {author} {\bibfnamefont {N.}~\bibnamefont {Asger~Mortensen}},\
  }\href {\doibase 10.1103/PhysRevLett.115.137403} {\bibfield  {journal}
  {\bibinfo  {journal} {Phys. Rev. Lett.}\ }\textbf {\bibinfo {volume} {115}},\
  \bibinfo {pages} {137403} (\bibinfo {year} {2015})}\BibitemShut {NoStop}%
\bibitem [{\citenamefont {Teperik}\ \emph {et~al.}(2016)\citenamefont
  {Teperik}, \citenamefont {Kazansky},\ and\ \citenamefont
  {Borisov}}]{Teperik2016}%
  \BibitemOpen
  \bibfield  {author} {\bibinfo {author} {\bibfnamefont {T.~V.}\ \bibnamefont
  {Teperik}}, \bibinfo {author} {\bibfnamefont {A.~K.}\ \bibnamefont
  {Kazansky}}, \ and\ \bibinfo {author} {\bibfnamefont {A.~G.}\ \bibnamefont
  {Borisov}},\ }\href {\doibase 10.1103/PhysRevB.93.155431} {\bibfield
  {journal} {\bibinfo  {journal} {Phys. Rev. B}\ }\textbf {\bibinfo {volume}
  {93}},\ \bibinfo {pages} {155431} (\bibinfo {year} {2016})}\BibitemShut
  {NoStop}%
\bibitem [{\citenamefont {Cirac\`{\i}}\ and\ \citenamefont
  {Della~Sala}(2016)}]{Ciraci2016}%
  \BibitemOpen
  \bibfield  {author} {\bibinfo {author} {\bibfnamefont {C.}~\bibnamefont
  {Cirac\`{\i}}}\ and\ \bibinfo {author} {\bibfnamefont {F.}~\bibnamefont
  {Della~Sala}},\ }\href {\doibase 10.1103/PhysRevB.93.205405} {\bibfield
  {journal} {\bibinfo  {journal} {Phys. Rev. B}\ }\textbf {\bibinfo {volume}
  {93}},\ \bibinfo {pages} {205405} (\bibinfo {year} {2016})}\BibitemShut
  {NoStop}%
\bibitem [{\citenamefont {David}\ \emph {et~al.}(2016)\citenamefont {David},
  \citenamefont {Christensen},\ and\ \citenamefont {Mortensen}}]{David2016}%
  \BibitemOpen
  \bibfield  {author} {\bibinfo {author} {\bibfnamefont {C.}~\bibnamefont
  {David}}, \bibinfo {author} {\bibfnamefont {J.}~\bibnamefont {Christensen}},
  \ and\ \bibinfo {author} {\bibfnamefont {N.~A.}\ \bibnamefont {Mortensen}},\
  }\href {\doibase 10.1103/PhysRevB.94.165410} {\bibfield  {journal} {\bibinfo
  {journal} {Phys. Rev. B}\ }\textbf {\bibinfo {volume} {94}},\ \bibinfo
  {pages} {165410} (\bibinfo {year} {2016})}\BibitemShut {NoStop}%
\bibitem [{\citenamefont {Christensen}\ \emph {et~al.}()\citenamefont
  {Christensen}, \citenamefont {Yan}, \citenamefont {Jauho}, \citenamefont
  {Soljacic},\ and\ \citenamefont {Mortensen}}]{Christensen2017}%
  \BibitemOpen
  \bibfield  {author} {\bibinfo {author} {\bibfnamefont {T.}~\bibnamefont
  {Christensen}}, \bibinfo {author} {\bibfnamefont {W.}~\bibnamefont {Yan}},
  \bibinfo {author} {\bibfnamefont {A.-P.}\ \bibnamefont {Jauho}}, \bibinfo
  {author} {\bibfnamefont {M.}~\bibnamefont {Soljacic}}, \ and\ \bibinfo
  {author} {\bibfnamefont {N.~A.}\ \bibnamefont {Mortensen}},\ }\href@noop {}
  {\ }\bibinfo {note} {Phys. Rev. Lett., in press (2017).
  \href{http://arXiv.org/abs/1608.05421v2}{arXiv:1608.05421v2}}\BibitemShut
  {NoStop}%
\bibitem [{\citenamefont {Cirac{\`\i}}\ \emph {et~al.}(2012)\citenamefont
  {Cirac{\`\i}}, \citenamefont {Hill}, \citenamefont {Mock}, \citenamefont
  {Urzhumov}, \citenamefont {Fern{\'a}ndez-Dom{\'\i}nguez}, \citenamefont
  {Maier}, \citenamefont {Pendry}, \citenamefont {Chilkoti},\ and\
  \citenamefont {Smith}}]{Ciraci2012}%
  \BibitemOpen
  \bibfield  {author} {\bibinfo {author} {\bibfnamefont {C.}~\bibnamefont
  {Cirac{\`\i}}}, \bibinfo {author} {\bibfnamefont {R.~T.}\ \bibnamefont
  {Hill}}, \bibinfo {author} {\bibfnamefont {J.~J.}\ \bibnamefont {Mock}},
  \bibinfo {author} {\bibfnamefont {Y.}~\bibnamefont {Urzhumov}}, \bibinfo
  {author} {\bibfnamefont {A.~I.}\ \bibnamefont
  {Fern{\'a}ndez-Dom{\'\i}nguez}}, \bibinfo {author} {\bibfnamefont {S.~A.}\
  \bibnamefont {Maier}}, \bibinfo {author} {\bibfnamefont {J.~B.}\ \bibnamefont
  {Pendry}}, \bibinfo {author} {\bibfnamefont {A.}~\bibnamefont {Chilkoti}}, \
  and\ \bibinfo {author} {\bibfnamefont {D.~R.}\ \bibnamefont {Smith}},\ }\href
  {\doibase 10.1126/science.1224823} {\bibfield  {journal} {\bibinfo  {journal}
  {Science}\ }\textbf {\bibinfo {volume} {337}},\ \bibinfo {pages} {1072}
  (\bibinfo {year} {2012})}\BibitemShut {NoStop}%
\bibitem [{\citenamefont {Scholl}\ \emph {et~al.}(2012)\citenamefont {Scholl},
  \citenamefont {Koh},\ and\ \citenamefont {Dionne}}]{Scholl2012}%
  \BibitemOpen
  \bibfield  {author} {\bibinfo {author} {\bibfnamefont {J.~A.}\ \bibnamefont
  {Scholl}}, \bibinfo {author} {\bibfnamefont {A.~L.}\ \bibnamefont {Koh}}, \
  and\ \bibinfo {author} {\bibfnamefont {J.~A.}\ \bibnamefont {Dionne}},\
  }\href {\doibase 10.1038/nature10904} {\bibfield  {journal} {\bibinfo
  {journal} {Nature}\ }\textbf {\bibinfo {volume} {483}},\ \bibinfo {pages}
  {421} (\bibinfo {year} {2012})}\BibitemShut {NoStop}%
\bibitem [{\citenamefont {Haberland}(2013)}]{Haberland2013}%
  \BibitemOpen
  \bibfield  {author} {\bibinfo {author} {\bibfnamefont {H.}~\bibnamefont
  {Haberland}},\ }\href {\doibase 10.1038/nature11886} {\bibfield  {journal}
  {\bibinfo  {journal} {Nature}\ }\textbf {\bibinfo {volume} {494}},\ \bibinfo
  {pages} {E1} (\bibinfo {year} {2013})}\BibitemShut {NoStop}%
\bibitem [{\citenamefont {Savage}\ \emph {et~al.}(2012)\citenamefont {Savage},
  \citenamefont {Hawkeye}, \citenamefont {Esteban}, \citenamefont {Borisov},
  \citenamefont {Aizpurua},\ and\ \citenamefont {Baumberg}}]{Savage2012}%
  \BibitemOpen
  \bibfield  {author} {\bibinfo {author} {\bibfnamefont {K.~J.}\ \bibnamefont
  {Savage}}, \bibinfo {author} {\bibfnamefont {M.~M.}\ \bibnamefont {Hawkeye}},
  \bibinfo {author} {\bibfnamefont {R.}~\bibnamefont {Esteban}}, \bibinfo
  {author} {\bibfnamefont {A.~G.}\ \bibnamefont {Borisov}}, \bibinfo {author}
  {\bibfnamefont {J.}~\bibnamefont {Aizpurua}}, \ and\ \bibinfo {author}
  {\bibfnamefont {J.~J.}\ \bibnamefont {Baumberg}},\ }\href {\doibase
  10.1038/nature11653} {\bibfield  {journal} {\bibinfo  {journal} {Nature}\
  }\textbf {\bibinfo {volume} {491}},\ \bibinfo {pages} {574} (\bibinfo {year}
  {2012})}\BibitemShut {NoStop}%
\bibitem [{\citenamefont {Banik}\ \emph {et~al.}(2012)\citenamefont {Banik},
  \citenamefont {El-Khoury}, \citenamefont {Nag}, \citenamefont
  {Rodriguez-Perez}, \citenamefont {Guarrottxena}, \citenamefont {Bazan},\ and\
  \citenamefont {Apkarian}}]{Banik2012}%
  \BibitemOpen
  \bibfield  {author} {\bibinfo {author} {\bibfnamefont {M.}~\bibnamefont
  {Banik}}, \bibinfo {author} {\bibfnamefont {P.~Z.}\ \bibnamefont
  {El-Khoury}}, \bibinfo {author} {\bibfnamefont {A.}~\bibnamefont {Nag}},
  \bibinfo {author} {\bibfnamefont {A.}~\bibnamefont {Rodriguez-Perez}},
  \bibinfo {author} {\bibfnamefont {N.}~\bibnamefont {Guarrottxena}}, \bibinfo
  {author} {\bibfnamefont {G.~C.}\ \bibnamefont {Bazan}}, \ and\ \bibinfo
  {author} {\bibfnamefont {V.~A.}\ \bibnamefont {Apkarian}},\ }\href {\doibase
  10.1021/nn304277n} {\bibfield  {journal} {\bibinfo  {journal} {ACS Nano}\
  }\textbf {\bibinfo {volume} {6}},\ \bibinfo {pages} {10343} (\bibinfo {year}
  {2012})}\BibitemShut {NoStop}%
\bibitem [{\citenamefont {Banik}\ \emph {et~al.}(2013)\citenamefont {Banik},
  \citenamefont {Apkarian}, \citenamefont {Park},\ and\ \citenamefont
  {Galperin}}]{Banik2013}%
  \BibitemOpen
  \bibfield  {author} {\bibinfo {author} {\bibfnamefont {M.}~\bibnamefont
  {Banik}}, \bibinfo {author} {\bibfnamefont {V.~A.}\ \bibnamefont {Apkarian}},
  \bibinfo {author} {\bibfnamefont {T.-H.}\ \bibnamefont {Park}}, \ and\
  \bibinfo {author} {\bibfnamefont {M.}~\bibnamefont {Galperin}},\ }\href
  {\doibase 10.1021/jz3018072} {\bibfield  {journal} {\bibinfo  {journal} {J.
  Phys. Chem. Lett.}\ }\textbf {\bibinfo {volume} {4}},\ \bibinfo {pages} {88}
  (\bibinfo {year} {2013})}\BibitemShut {NoStop}%
\bibitem [{\citenamefont {Scholl}\ \emph {et~al.}(2013)\citenamefont {Scholl},
  \citenamefont {Garc{\'i}a-Etxarri}, \citenamefont {Koh},\ and\ \citenamefont
  {Dionne}}]{Scholl2013}%
  \BibitemOpen
  \bibfield  {author} {\bibinfo {author} {\bibfnamefont {J.~A.}\ \bibnamefont
  {Scholl}}, \bibinfo {author} {\bibfnamefont {A.}~\bibnamefont
  {Garc{\'i}a-Etxarri}}, \bibinfo {author} {\bibfnamefont {A.~L.}\ \bibnamefont
  {Koh}}, \ and\ \bibinfo {author} {\bibfnamefont {J.~A.}\ \bibnamefont
  {Dionne}},\ }\href {\doibase 10.1021/nl304078v} {\bibfield  {journal}
  {\bibinfo  {journal} {Nano Lett.}\ }\textbf {\bibinfo {volume} {13}},\
  \bibinfo {pages} {564} (\bibinfo {year} {2013})}\BibitemShut {NoStop}%
\bibitem [{\citenamefont {Tan}\ \emph {et~al.}(2014)\citenamefont {Tan},
  \citenamefont {Wu}, \citenamefont {Yang}, \citenamefont {Bai}, \citenamefont
  {Bosman},\ and\ \citenamefont {Nijhuis}}]{Tan2014}%
  \BibitemOpen
  \bibfield  {author} {\bibinfo {author} {\bibfnamefont {S.~F.}\ \bibnamefont
  {Tan}}, \bibinfo {author} {\bibfnamefont {L.}~\bibnamefont {Wu}}, \bibinfo
  {author} {\bibfnamefont {J.~K.}\ \bibnamefont {Yang}}, \bibinfo {author}
  {\bibfnamefont {P.}~\bibnamefont {Bai}}, \bibinfo {author} {\bibfnamefont
  {M.}~\bibnamefont {Bosman}}, \ and\ \bibinfo {author} {\bibfnamefont {C.~A.}\
  \bibnamefont {Nijhuis}},\ }\href {\doibase 10.1126/science.1248797}
  {\bibfield  {journal} {\bibinfo  {journal} {Science}\ }\textbf {\bibinfo
  {volume} {343}},\ \bibinfo {pages} {1496} (\bibinfo {year}
  {2014})}\BibitemShut {NoStop}%
\bibitem [{\citenamefont {Zhang}\ \emph {et~al.}(2016)\citenamefont {Zhang},
  \citenamefont {Zhao}, \citenamefont {Zhou}, \citenamefont {Schlather},
  \citenamefont {Dong}, \citenamefont {McClain}, \citenamefont {Swearer},
  \citenamefont {Nordlander},\ and\ \citenamefont {Halas}}]{Zhang2016}%
  \BibitemOpen
  \bibfield  {author} {\bibinfo {author} {\bibfnamefont {C.}~\bibnamefont
  {Zhang}}, \bibinfo {author} {\bibfnamefont {H.}~\bibnamefont {Zhao}},
  \bibinfo {author} {\bibfnamefont {L.}~\bibnamefont {Zhou}}, \bibinfo {author}
  {\bibfnamefont {A.~E.}\ \bibnamefont {Schlather}}, \bibinfo {author}
  {\bibfnamefont {L.}~\bibnamefont {Dong}}, \bibinfo {author} {\bibfnamefont
  {M.~J.}\ \bibnamefont {McClain}}, \bibinfo {author} {\bibfnamefont {D.~F.}\
  \bibnamefont {Swearer}}, \bibinfo {author} {\bibfnamefont {P.}~\bibnamefont
  {Nordlander}}, \ and\ \bibinfo {author} {\bibfnamefont {N.~J.}\ \bibnamefont
  {Halas}},\ }\href {\doibase 10.1021/acs.nanolett.6b03582} {\bibfield
  {journal} {\bibinfo  {journal} {Nano Lett.}\ }\textbf {\bibinfo {volume}
  {16}},\ \bibinfo {pages} {6677} (\bibinfo {year} {2016})}\BibitemShut
  {NoStop}%
\bibitem [{\citenamefont {Sanders}\ \emph {et~al.}(2016)\citenamefont
  {Sanders}, \citenamefont {Bowman},\ and\ \citenamefont
  {Baumberg}}]{Sanders2016}%
  \BibitemOpen
  \bibfield  {author} {\bibinfo {author} {\bibfnamefont {A.}~\bibnamefont
  {Sanders}}, \bibinfo {author} {\bibfnamefont {R.~W.}\ \bibnamefont {Bowman}},
  \ and\ \bibinfo {author} {\bibfnamefont {J.~J.}\ \bibnamefont {Baumberg}},\
  }\href {\doibase 10.1038/srep32988} {\bibfield  {journal} {\bibinfo
  {journal} {Sci. Rep.}\ }\textbf {\bibinfo {volume} {6}},\ \bibinfo {pages}
  {32988} (\bibinfo {year} {2016})}\BibitemShut {NoStop}%
\bibitem [{\citenamefont {Mertens}\ \emph {et~al.}(2016)\citenamefont
  {Mertens}, \citenamefont {Demetriadou}, \citenamefont {Bowman}, \citenamefont
  {Benz}, \citenamefont {Kleemann}, \citenamefont {Tserkezis}, \citenamefont
  {Shi}, \citenamefont {Yang}, \citenamefont {Hess}, \citenamefont {Aizpurua},\
  and\ \citenamefont {Baumberg}}]{Mertens2016}%
  \BibitemOpen
  \bibfield  {author} {\bibinfo {author} {\bibfnamefont {J.}~\bibnamefont
  {Mertens}}, \bibinfo {author} {\bibfnamefont {A.}~\bibnamefont
  {Demetriadou}}, \bibinfo {author} {\bibfnamefont {R.~W.}\ \bibnamefont
  {Bowman}}, \bibinfo {author} {\bibfnamefont {F.}~\bibnamefont {Benz}},
  \bibinfo {author} {\bibfnamefont {M.-E.}\ \bibnamefont {Kleemann}}, \bibinfo
  {author} {\bibfnamefont {C.}~\bibnamefont {Tserkezis}}, \bibinfo {author}
  {\bibfnamefont {Y.}~\bibnamefont {Shi}}, \bibinfo {author} {\bibfnamefont
  {H.~Y.}\ \bibnamefont {Yang}}, \bibinfo {author} {\bibfnamefont
  {O.}~\bibnamefont {Hess}}, \bibinfo {author} {\bibfnamefont {J.}~\bibnamefont
  {Aizpurua}}, \ and\ \bibinfo {author} {\bibfnamefont {J.~J.}\ \bibnamefont
  {Baumberg}},\ }\href {\doibase 10.1021/acs.nanolett.6b02164} {\bibfield
  {journal} {\bibinfo  {journal} {Nano Lett.}\ }\textbf {\bibinfo {volume}
  {16}},\ \bibinfo {pages} {5605} (\bibinfo {year} {2016})}\BibitemShut
  {NoStop}%
\bibitem [{\citenamefont {Tame}\ \emph {et~al.}(2013)\citenamefont {Tame},
  \citenamefont {McEnery}, \citenamefont {Ozdemir}, \citenamefont {Lee},
  \citenamefont {Maier},\ and\ \citenamefont {Kim}}]{Tame2013}%
  \BibitemOpen
  \bibfield  {author} {\bibinfo {author} {\bibfnamefont {M.~S.}\ \bibnamefont
  {Tame}}, \bibinfo {author} {\bibfnamefont {K.~R.}\ \bibnamefont {McEnery}},
  \bibinfo {author} {\bibfnamefont {S.~K.}\ \bibnamefont {Ozdemir}}, \bibinfo
  {author} {\bibfnamefont {J.}~\bibnamefont {Lee}}, \bibinfo {author}
  {\bibfnamefont {S.~A.}\ \bibnamefont {Maier}}, \ and\ \bibinfo {author}
  {\bibfnamefont {M.~S.}\ \bibnamefont {Kim}},\ }\href {\doibase
  10.1038/nphys2615} {\bibfield  {journal} {\bibinfo  {journal} {Nat. Phys.}\
  }\textbf {\bibinfo {volume} {9}},\ \bibinfo {pages} {329} (\bibinfo {year}
  {2013})}\BibitemShut {NoStop}%
\bibitem [{\citenamefont {Zhu}\ \emph {et~al.}(2016)\citenamefont {Zhu},
  \citenamefont {Esteban}, \citenamefont {Borisov}, \citenamefont {Baumberg},
  \citenamefont {Nordlander}, \citenamefont {Lezec}, \citenamefont {Aizpurua},\
  and\ \citenamefont {Crozier}}]{Zhu2016}%
  \BibitemOpen
  \bibfield  {author} {\bibinfo {author} {\bibfnamefont {W.}~\bibnamefont
  {Zhu}}, \bibinfo {author} {\bibfnamefont {R.}~\bibnamefont {Esteban}},
  \bibinfo {author} {\bibfnamefont {A.~G.}\ \bibnamefont {Borisov}}, \bibinfo
  {author} {\bibfnamefont {J.~J.}\ \bibnamefont {Baumberg}}, \bibinfo {author}
  {\bibfnamefont {P.}~\bibnamefont {Nordlander}}, \bibinfo {author}
  {\bibfnamefont {H.~J.}\ \bibnamefont {Lezec}}, \bibinfo {author}
  {\bibfnamefont {J.}~\bibnamefont {Aizpurua}}, \ and\ \bibinfo {author}
  {\bibfnamefont {K.~B.}\ \bibnamefont {Crozier}},\ }\href {\doibase
  10.1038/ncomms11495} {\bibfield  {journal} {\bibinfo  {journal} {Nat.
  Commun.}\ }\textbf {\bibinfo {volume} {7}},\ \bibinfo {pages} {11495}
  (\bibinfo {year} {2016})}\BibitemShut {NoStop}%
\bibitem [{\citenamefont {Iida}\ \emph {et~al.}(2014)\citenamefont {Iida},
  \citenamefont {Noda}, \citenamefont {Ishimura},\ and\ \citenamefont
  {Nobusada}}]{Iida2014}%
  \BibitemOpen
  \bibfield  {author} {\bibinfo {author} {\bibfnamefont {K.}~\bibnamefont
  {Iida}}, \bibinfo {author} {\bibfnamefont {M.}~\bibnamefont {Noda}}, \bibinfo
  {author} {\bibfnamefont {K.}~\bibnamefont {Ishimura}}, \ and\ \bibinfo
  {author} {\bibfnamefont {K.}~\bibnamefont {Nobusada}},\ }\href {\doibase
  10.1021/jp5088042} {\bibfield  {journal} {\bibinfo  {journal} {J. Phys. Chem.
  A}\ }\textbf {\bibinfo {volume} {118}},\ \bibinfo {pages} {11317} (\bibinfo
  {year} {2014})}\BibitemShut {NoStop}%
\bibitem [{\citenamefont {Kuisma}\ \emph {et~al.}(2015)\citenamefont {Kuisma},
  \citenamefont {Sakko}, \citenamefont {Rossi}, \citenamefont {Larsen},
  \citenamefont {Enkovaara}, \citenamefont {Lehtovaara},\ and\ \citenamefont
  {Rantala}}]{Kuisma2015}%
  \BibitemOpen
  \bibfield  {author} {\bibinfo {author} {\bibfnamefont {M.}~\bibnamefont
  {Kuisma}}, \bibinfo {author} {\bibfnamefont {A.}~\bibnamefont {Sakko}},
  \bibinfo {author} {\bibfnamefont {T.~P.}\ \bibnamefont {Rossi}}, \bibinfo
  {author} {\bibfnamefont {A.~H.}\ \bibnamefont {Larsen}}, \bibinfo {author}
  {\bibfnamefont {J.}~\bibnamefont {Enkovaara}}, \bibinfo {author}
  {\bibfnamefont {L.}~\bibnamefont {Lehtovaara}}, \ and\ \bibinfo {author}
  {\bibfnamefont {T.~T.}\ \bibnamefont {Rantala}},\ }\href {\doibase
  10.1103/PhysRevB.91.115431} {\bibfield  {journal} {\bibinfo  {journal} {Phys.
  Rev. B}\ }\textbf {\bibinfo {volume} {91}},\ \bibinfo {pages} {115431}
  (\bibinfo {year} {2015})}\BibitemShut {NoStop}%
\bibitem [{\citenamefont {Baseggio}\ \emph {et~al.}(2015)\citenamefont
  {Baseggio}, \citenamefont {Fronzoni},\ and\ \citenamefont
  {Stener}}]{Baseggio2015}%
  \BibitemOpen
  \bibfield  {author} {\bibinfo {author} {\bibfnamefont {O.}~\bibnamefont
  {Baseggio}}, \bibinfo {author} {\bibfnamefont {G.}~\bibnamefont {Fronzoni}},
  \ and\ \bibinfo {author} {\bibfnamefont {M.}~\bibnamefont {Stener}},\ }\href
  {\doibase 10.1063/1.4923368} {\bibfield  {journal} {\bibinfo  {journal} {J.
  Chem. Phys.}\ }\textbf {\bibinfo {volume} {143}},\ \bibinfo {pages} {024106}
  (\bibinfo {year} {2015})}\BibitemShut {NoStop}%
\bibitem [{\citenamefont {Baseggio}\ \emph {et~al.}(2016)\citenamefont
  {Baseggio}, \citenamefont {De~Vetta}, \citenamefont {Fronzoni}, \citenamefont
  {Stener}, \citenamefont {Sementa}, \citenamefont {Fortunelli},\ and\
  \citenamefont {Calzolari}}]{Baseggio2016}%
  \BibitemOpen
  \bibfield  {author} {\bibinfo {author} {\bibfnamefont {O.}~\bibnamefont
  {Baseggio}}, \bibinfo {author} {\bibfnamefont {M.}~\bibnamefont {De~Vetta}},
  \bibinfo {author} {\bibfnamefont {G.}~\bibnamefont {Fronzoni}}, \bibinfo
  {author} {\bibfnamefont {M.}~\bibnamefont {Stener}}, \bibinfo {author}
  {\bibfnamefont {L.}~\bibnamefont {Sementa}}, \bibinfo {author} {\bibfnamefont
  {A.}~\bibnamefont {Fortunelli}}, \ and\ \bibinfo {author} {\bibfnamefont
  {A.}~\bibnamefont {Calzolari}},\ }\href {\doibase 10.1021/acs.jpcc.6b04709}
  {\bibfield  {journal} {\bibinfo  {journal} {J. Phys. Chem. C}\ }\textbf
  {\bibinfo {volume} {120}},\ \bibinfo {pages} {12773} (\bibinfo {year}
  {2016})}\BibitemShut {NoStop}%
\bibitem [{\citenamefont {Koval}\ \emph {et~al.}(2016)\citenamefont {Koval},
  \citenamefont {Marchesin}, \citenamefont {Foerster},\ and\ \citenamefont
  {S{\'a}nchez-Portal}}]{Koval2016}%
  \BibitemOpen
  \bibfield  {author} {\bibinfo {author} {\bibfnamefont {P.}~\bibnamefont
  {Koval}}, \bibinfo {author} {\bibfnamefont {F.}~\bibnamefont {Marchesin}},
  \bibinfo {author} {\bibfnamefont {D.}~\bibnamefont {Foerster}}, \ and\
  \bibinfo {author} {\bibfnamefont {D.}~\bibnamefont {S{\'a}nchez-Portal}},\
  }\href {\doibase 10.1088/0953-8984/28/21/214001} {\bibfield  {journal}
  {\bibinfo  {journal} {J. Phys.: Condens. Matter}\ }\textbf {\bibinfo {volume}
  {28}},\ \bibinfo {pages} {214001} (\bibinfo {year} {2016})}\BibitemShut
  {NoStop}%
\bibitem [{\citenamefont {Casida}(1995)}]{Casida1995}%
  \BibitemOpen
  \bibfield  {author} {\bibinfo {author} {\bibfnamefont {M.~E.}\ \bibnamefont
  {Casida}},\ }in\ \href {\doibase 10.1142/9789812830586_0005} {\emph {\bibinfo
  {booktitle} {{Recent Advances in Density Functional Methods, Part I}}}},\
  \bibinfo {editor} {edited by\ \bibinfo {editor} {\bibfnamefont {D.~P.}\
  \bibnamefont {Chong}}}\ (\bibinfo  {publisher} {World Scientific,
  Singapore},\ \bibinfo {year} {1995})\ p.\ \bibinfo {pages} {155}\BibitemShut
  {NoStop}%
\bibitem [{\citenamefont {Petersilka}\ \emph {et~al.}(1996)\citenamefont
  {Petersilka}, \citenamefont {Gossmann},\ and\ \citenamefont
  {Gross}}]{Petersilka1996}%
  \BibitemOpen
  \bibfield  {author} {\bibinfo {author} {\bibfnamefont {M.}~\bibnamefont
  {Petersilka}}, \bibinfo {author} {\bibfnamefont {U.~J.}\ \bibnamefont
  {Gossmann}}, \ and\ \bibinfo {author} {\bibfnamefont {E.~K.~U.}\ \bibnamefont
  {Gross}},\ }\href {\doibase 10.1103/PhysRevLett.76.1212} {\bibfield
  {journal} {\bibinfo  {journal} {Phys. Rev. Lett.}\ }\textbf {\bibinfo
  {volume} {76}},\ \bibinfo {pages} {1212} (\bibinfo {year}
  {1996})}\BibitemShut {NoStop}%
\bibitem [{\citenamefont {Casida}(2009)}]{Casida2009}%
  \BibitemOpen
  \bibfield  {author} {\bibinfo {author} {\bibfnamefont {M.~E.}\ \bibnamefont
  {Casida}},\ }\href {\doibase 10.1016/j.theochem.2009.08.018} {\bibfield
  {journal} {\bibinfo  {journal} {J. Mol. Struct. {THEOCHEM}}\ }\textbf
  {\bibinfo {volume} {914}},\ \bibinfo {pages} {3 } (\bibinfo {year}
  {2009})}\BibitemShut {NoStop}%
\bibitem [{\citenamefont {Bauernschmitt}\ and\ \citenamefont
  {Ahlrichs}(1996)}]{Bauernschmitt1996}%
  \BibitemOpen
  \bibfield  {author} {\bibinfo {author} {\bibfnamefont {R.}~\bibnamefont
  {Bauernschmitt}}\ and\ \bibinfo {author} {\bibfnamefont {R.}~\bibnamefont
  {Ahlrichs}},\ }\href {\doibase 10.1016/0009-2614(96)00440-X} {\bibfield
  {journal} {\bibinfo  {journal} {Chem. Phys. Lett.}\ }\textbf {\bibinfo
  {volume} {256}},\ \bibinfo {pages} {454 } (\bibinfo {year}
  {1996})}\BibitemShut {NoStop}%
\bibitem [{\citenamefont {Stratmann}\ \emph {et~al.}(1998)\citenamefont
  {Stratmann}, \citenamefont {Scuseria},\ and\ \citenamefont
  {Frisch}}]{Stratmann1998}%
  \BibitemOpen
  \bibfield  {author} {\bibinfo {author} {\bibfnamefont {R.~E.}\ \bibnamefont
  {Stratmann}}, \bibinfo {author} {\bibfnamefont {G.~E.}\ \bibnamefont
  {Scuseria}}, \ and\ \bibinfo {author} {\bibfnamefont {M.~J.}\ \bibnamefont
  {Frisch}},\ }\href {\doibase 10.1063/1.477483} {\bibfield  {journal}
  {\bibinfo  {journal} {J. Chem. Phys.}\ }\textbf {\bibinfo {volume} {109}},\
  \bibinfo {pages} {8218} (\bibinfo {year} {1998})}\BibitemShut {NoStop}%
\bibitem [{\citenamefont {Walker}\ \emph {et~al.}(2006)\citenamefont {Walker},
  \citenamefont {Saitta}, \citenamefont {Gebauer},\ and\ \citenamefont
  {Baroni}}]{Walker2006}%
  \BibitemOpen
  \bibfield  {author} {\bibinfo {author} {\bibfnamefont {B.}~\bibnamefont
  {Walker}}, \bibinfo {author} {\bibfnamefont {A.~M.}\ \bibnamefont {Saitta}},
  \bibinfo {author} {\bibfnamefont {R.}~\bibnamefont {Gebauer}}, \ and\
  \bibinfo {author} {\bibfnamefont {S.}~\bibnamefont {Baroni}},\ }\href
  {\doibase 10.1103/PhysRevLett.96.113001} {\bibfield  {journal} {\bibinfo
  {journal} {Phys. Rev. Lett.}\ }\textbf {\bibinfo {volume} {96}},\ \bibinfo
  {pages} {113001} (\bibinfo {year} {2006})}\BibitemShut {NoStop}%
\bibitem [{\citenamefont {Andrade}\ \emph {et~al.}(2007)\citenamefont
  {Andrade}, \citenamefont {Botti}, \citenamefont {Marques},\ and\
  \citenamefont {Rubio}}]{Andrade2007}%
  \BibitemOpen
  \bibfield  {author} {\bibinfo {author} {\bibfnamefont {X.}~\bibnamefont
  {Andrade}}, \bibinfo {author} {\bibfnamefont {S.}~\bibnamefont {Botti}},
  \bibinfo {author} {\bibfnamefont {M.~A.~L.}\ \bibnamefont {Marques}}, \ and\
  \bibinfo {author} {\bibfnamefont {A.}~\bibnamefont {Rubio}},\ }\href
  {\doibase 10.1063/1.2733666} {\bibfield  {journal} {\bibinfo  {journal} {J.
  Chem. Phys.}\ }\textbf {\bibinfo {volume} {126}},\ \bibinfo {pages} {184106}
  (\bibinfo {year} {2007})}\BibitemShut {NoStop}%
\bibitem [{\citenamefont {Yabana}\ and\ \citenamefont
  {Bertsch}(1996)}]{Yabana1996}%
  \BibitemOpen
  \bibfield  {author} {\bibinfo {author} {\bibfnamefont {K.}~\bibnamefont
  {Yabana}}\ and\ \bibinfo {author} {\bibfnamefont {G.~F.}\ \bibnamefont
  {Bertsch}},\ }\href {\doibase 10.1103/PhysRevB.54.4484} {\bibfield  {journal}
  {\bibinfo  {journal} {Phys. Rev. B}\ }\textbf {\bibinfo {volume} {54}},\
  \bibinfo {pages} {4484} (\bibinfo {year} {1996})}\BibitemShut {NoStop}%
\bibitem [{\citenamefont {Yabana}\ \emph {et~al.}(2006)\citenamefont {Yabana},
  \citenamefont {Nakatsukasa}, \citenamefont {Iwata},\ and\ \citenamefont
  {Bertsch}}]{Yabana2006}%
  \BibitemOpen
  \bibfield  {author} {\bibinfo {author} {\bibfnamefont {K.}~\bibnamefont
  {Yabana}}, \bibinfo {author} {\bibfnamefont {T.}~\bibnamefont {Nakatsukasa}},
  \bibinfo {author} {\bibfnamefont {J.-I.}\ \bibnamefont {Iwata}}, \ and\
  \bibinfo {author} {\bibfnamefont {G.~F.}\ \bibnamefont {Bertsch}},\ }\href
  {\doibase 10.1002/pssb.200642005} {\bibfield  {journal} {\bibinfo  {journal}
  {Phys. Status Solidi B}\ }\textbf {\bibinfo {volume} {243}},\ \bibinfo
  {pages} {1121} (\bibinfo {year} {2006})}\BibitemShut {NoStop}%
\bibitem [{\citenamefont {Sander}\ and\ \citenamefont
  {Kresse}(2017)}]{Sander2017}%
  \BibitemOpen
  \bibfield  {author} {\bibinfo {author} {\bibfnamefont {T.}~\bibnamefont
  {Sander}}\ and\ \bibinfo {author} {\bibfnamefont {G.}~\bibnamefont
  {Kresse}},\ }\href {\doibase 10.1063/1.4975193} {\bibfield  {journal}
  {\bibinfo  {journal} {J. Chem. Phys.}\ }\textbf {\bibinfo {volume} {146}},\
  \bibinfo {pages} {064110} (\bibinfo {year} {2017})}\BibitemShut {NoStop}%
\bibitem [{\citenamefont {Yan}\ \emph {et~al.}(2016)\citenamefont {Yan},
  \citenamefont {Wang},\ and\ \citenamefont {Meng}}]{Yan2016}%
  \BibitemOpen
  \bibfield  {author} {\bibinfo {author} {\bibfnamefont {L.}~\bibnamefont
  {Yan}}, \bibinfo {author} {\bibfnamefont {F.}~\bibnamefont {Wang}}, \ and\
  \bibinfo {author} {\bibfnamefont {S.}~\bibnamefont {Meng}},\ }\href {\doibase
  10.1021/acsnano.6b01840} {\bibfield  {journal} {\bibinfo  {journal} {ACS
  Nano}\ }\textbf {\bibinfo {volume} {10}},\ \bibinfo {pages} {5452} (\bibinfo
  {year} {2016})}\BibitemShut {NoStop}%
\bibitem [{\citenamefont {Li}\ and\ \citenamefont {Ullrich}(2011)}]{Li2011}%
  \BibitemOpen
  \bibfield  {author} {\bibinfo {author} {\bibfnamefont {Y.}~\bibnamefont
  {Li}}\ and\ \bibinfo {author} {\bibfnamefont {C.~A.}\ \bibnamefont
  {Ullrich}},\ }\href {\doibase 10.1016/j.chemphys.2011.02.001} {\bibfield
  {journal} {\bibinfo  {journal} {Chem. Phys.}\ }\textbf {\bibinfo {volume}
  {391}},\ \bibinfo {pages} {157 } (\bibinfo {year} {2011})}\BibitemShut
  {NoStop}%
\bibitem [{\citenamefont {Li}\ and\ \citenamefont {Ullrich}(2015)}]{Li2015}%
  \BibitemOpen
  \bibfield  {author} {\bibinfo {author} {\bibfnamefont {Y.}~\bibnamefont
  {Li}}\ and\ \bibinfo {author} {\bibfnamefont {C.~A.}\ \bibnamefont
  {Ullrich}},\ }\href {\doibase 10.1021/acs.jctc.5b00987} {\bibfield  {journal}
  {\bibinfo  {journal} {J. Chem. Theory Comput.}\ }\textbf {\bibinfo {volume}
  {11}},\ \bibinfo {pages} {5838} (\bibinfo {year} {2015})}\BibitemShut
  {NoStop}%
\bibitem [{\citenamefont {Repisky}\ \emph {et~al.}(2015)\citenamefont
  {Repisky}, \citenamefont {Konecny}, \citenamefont {Kadek}, \citenamefont
  {Komorovsky}, \citenamefont {Malkin}, \citenamefont {Malkin},\ and\
  \citenamefont {Ruud}}]{Repisky2015}%
  \BibitemOpen
  \bibfield  {author} {\bibinfo {author} {\bibfnamefont {M.}~\bibnamefont
  {Repisky}}, \bibinfo {author} {\bibfnamefont {L.}~\bibnamefont {Konecny}},
  \bibinfo {author} {\bibfnamefont {M.}~\bibnamefont {Kadek}}, \bibinfo
  {author} {\bibfnamefont {S.}~\bibnamefont {Komorovsky}}, \bibinfo {author}
  {\bibfnamefont {O.~L.}\ \bibnamefont {Malkin}}, \bibinfo {author}
  {\bibfnamefont {V.~G.}\ \bibnamefont {Malkin}}, \ and\ \bibinfo {author}
  {\bibfnamefont {K.}~\bibnamefont {Ruud}},\ }\href {\doibase
  10.1021/ct501078d} {\bibfield  {journal} {\bibinfo  {journal} {J. Chem.
  Theory Comput.}\ }\textbf {\bibinfo {volume} {11}},\ \bibinfo {pages} {980}
  (\bibinfo {year} {2015})}\BibitemShut {NoStop}%
\bibitem [{\citenamefont {Kolesov}\ \emph {et~al.}(2016)\citenamefont
  {Kolesov}, \citenamefont {Gr\r{a}n{\"a}s}, \citenamefont {Hoyt},
  \citenamefont {Vinichenko},\ and\ \citenamefont {Kaxiras}}]{Kolesov2016}%
  \BibitemOpen
  \bibfield  {author} {\bibinfo {author} {\bibfnamefont {G.}~\bibnamefont
  {Kolesov}}, \bibinfo {author} {\bibfnamefont {O.}~\bibnamefont
  {Gr\r{a}n{\"a}s}}, \bibinfo {author} {\bibfnamefont {R.}~\bibnamefont
  {Hoyt}}, \bibinfo {author} {\bibfnamefont {D.}~\bibnamefont {Vinichenko}}, \
  and\ \bibinfo {author} {\bibfnamefont {E.}~\bibnamefont {Kaxiras}},\ }\href
  {\doibase 10.1021/acs.jctc.5b00969} {\bibfield  {journal} {\bibinfo
  {journal} {J. Chem. Theory Comput.}\ }\textbf {\bibinfo {volume} {12}},\
  \bibinfo {pages} {466} (\bibinfo {year} {2016})}\BibitemShut {NoStop}%
\bibitem [{\citenamefont {Malola}\ \emph {et~al.}(2014)\citenamefont {Malola},
  \citenamefont {Lehtovaara},\ and\ \citenamefont {H{\"a}kkinen}}]{Malola2014}%
  \BibitemOpen
  \bibfield  {author} {\bibinfo {author} {\bibfnamefont {S.}~\bibnamefont
  {Malola}}, \bibinfo {author} {\bibfnamefont {L.}~\bibnamefont {Lehtovaara}},
  \ and\ \bibinfo {author} {\bibfnamefont {H.}~\bibnamefont {H{\"a}kkinen}},\
  }\href {\doibase 10.1021/jp505462m} {\bibfield  {journal} {\bibinfo
  {journal} {J. Phys. Chem. C}\ }\textbf {\bibinfo {volume} {118}},\ \bibinfo
  {pages} {20002} (\bibinfo {year} {2014})}\BibitemShut {NoStop}%
\bibitem [{\citenamefont {Malola}\ \emph {et~al.}(2015)\citenamefont {Malola},
  \citenamefont {Hartmann},\ and\ \citenamefont {H{\"a}kkinen}}]{Malola2015}%
  \BibitemOpen
  \bibfield  {author} {\bibinfo {author} {\bibfnamefont {S.}~\bibnamefont
  {Malola}}, \bibinfo {author} {\bibfnamefont {M.~J.}\ \bibnamefont
  {Hartmann}}, \ and\ \bibinfo {author} {\bibfnamefont {H.}~\bibnamefont
  {H{\"a}kkinen}},\ }\href {\doibase 10.1021/jz502637b} {\bibfield  {journal}
  {\bibinfo  {journal} {J. Phys. Chem. Lett.}\ }\textbf {\bibinfo {volume}
  {6}},\ \bibinfo {pages} {515} (\bibinfo {year} {2015})}\BibitemShut {NoStop}%
\bibitem [{\citenamefont {Larsen}\ \emph {et~al.}(2009)\citenamefont {Larsen},
  \citenamefont {Vanin}, \citenamefont {Mortensen}, \citenamefont {Thygesen},\
  and\ \citenamefont {Jacobsen}}]{Larsen2009}%
  \BibitemOpen
  \bibfield  {author} {\bibinfo {author} {\bibfnamefont {A.~H.}\ \bibnamefont
  {Larsen}}, \bibinfo {author} {\bibfnamefont {M.}~\bibnamefont {Vanin}},
  \bibinfo {author} {\bibfnamefont {J.~J.}\ \bibnamefont {Mortensen}}, \bibinfo
  {author} {\bibfnamefont {K.~S.}\ \bibnamefont {Thygesen}}, \ and\ \bibinfo
  {author} {\bibfnamefont {K.~W.}\ \bibnamefont {Jacobsen}},\ }\href {\doibase
  10.1103/PhysRevB.80.195112} {\bibfield  {journal} {\bibinfo  {journal} {Phys.
  Rev. B}\ }\textbf {\bibinfo {volume} {80}},\ \bibinfo {pages} {195112}
  (\bibinfo {year} {2009})}\BibitemShut {NoStop}%
\bibitem [{\citenamefont {Mortensen}\ \emph {et~al.}(2005)\citenamefont
  {Mortensen}, \citenamefont {Hansen},\ and\ \citenamefont
  {Jacobsen}}]{Mortensen2005}%
  \BibitemOpen
  \bibfield  {author} {\bibinfo {author} {\bibfnamefont {J.~J.}\ \bibnamefont
  {Mortensen}}, \bibinfo {author} {\bibfnamefont {L.~B.}\ \bibnamefont
  {Hansen}}, \ and\ \bibinfo {author} {\bibfnamefont {K.~W.}\ \bibnamefont
  {Jacobsen}},\ }\href {\doibase 10.1103/PhysRevB.71.035109} {\bibfield
  {journal} {\bibinfo  {journal} {Phys. Rev. B}\ }\textbf {\bibinfo {volume}
  {71}},\ \bibinfo {pages} {035109} (\bibinfo {year} {2005})}\BibitemShut
  {NoStop}%
\bibitem [{\citenamefont {Enkovaara}\ \emph {et~al.}(2010)\citenamefont
  {Enkovaara}, \citenamefont {Rostgaard}, \citenamefont {Mortensen},
  \citenamefont {Chen}, \citenamefont {Dułak}, \citenamefont {Ferrighi},
  \citenamefont {Gavnholt}, \citenamefont {Glinsvad}, \citenamefont {Haikola},
  \citenamefont {Hansen}, \citenamefont {Kristoffersen}, \citenamefont
  {Kuisma}, \citenamefont {Larsen}, \citenamefont {Lehtovaara}, \citenamefont
  {Ljungberg}, \citenamefont {Lopez-Acevedo}, \citenamefont {Moses},
  \citenamefont {Ojanen}, \citenamefont {Olsen}, \citenamefont {Petzold},
  \citenamefont {Romero}, \citenamefont {Stausholm-M{\o}ller}, \citenamefont
  {Strange}, \citenamefont {Tritsaris}, \citenamefont {Vanin}, \citenamefont
  {Walter}, \citenamefont {Hammer}, \citenamefont {H{\"a}kkinen}, \citenamefont
  {Madsen}, \citenamefont {Nieminen}, \citenamefont {N{\o}rskov}, \citenamefont
  {Puska}, \citenamefont {Rantala}, \citenamefont {Schi{\o}tz}, \citenamefont
  {Thygesen},\ and\ \citenamefont {Jacobsen}}]{Enkovaara2010}%
  \BibitemOpen
  \bibfield  {author} {\bibinfo {author} {\bibfnamefont {J.}~\bibnamefont
  {Enkovaara}}, \bibinfo {author} {\bibfnamefont {C.}~\bibnamefont
  {Rostgaard}}, \bibinfo {author} {\bibfnamefont {J.~J.}\ \bibnamefont
  {Mortensen}}, \bibinfo {author} {\bibfnamefont {J.}~\bibnamefont {Chen}},
  \bibinfo {author} {\bibfnamefont {M.}~\bibnamefont {Dułak}}, \bibinfo
  {author} {\bibfnamefont {L.}~\bibnamefont {Ferrighi}}, \bibinfo {author}
  {\bibfnamefont {J.}~\bibnamefont {Gavnholt}}, \bibinfo {author}
  {\bibfnamefont {C.}~\bibnamefont {Glinsvad}}, \bibinfo {author}
  {\bibfnamefont {V.}~\bibnamefont {Haikola}}, \bibinfo {author} {\bibfnamefont
  {H.~A.}\ \bibnamefont {Hansen}}, \bibinfo {author} {\bibfnamefont {H.~H.}\
  \bibnamefont {Kristoffersen}}, \bibinfo {author} {\bibfnamefont
  {M.}~\bibnamefont {Kuisma}}, \bibinfo {author} {\bibfnamefont {A.~H.}\
  \bibnamefont {Larsen}}, \bibinfo {author} {\bibfnamefont {L.}~\bibnamefont
  {Lehtovaara}}, \bibinfo {author} {\bibfnamefont {M.}~\bibnamefont
  {Ljungberg}}, \bibinfo {author} {\bibfnamefont {O.}~\bibnamefont
  {Lopez-Acevedo}}, \bibinfo {author} {\bibfnamefont {P.~G.}\ \bibnamefont
  {Moses}}, \bibinfo {author} {\bibfnamefont {J.}~\bibnamefont {Ojanen}},
  \bibinfo {author} {\bibfnamefont {T.}~\bibnamefont {Olsen}}, \bibinfo
  {author} {\bibfnamefont {V.}~\bibnamefont {Petzold}}, \bibinfo {author}
  {\bibfnamefont {N.~A.}\ \bibnamefont {Romero}}, \bibinfo {author}
  {\bibfnamefont {J.}~\bibnamefont {Stausholm-M{\o}ller}}, \bibinfo {author}
  {\bibfnamefont {M.}~\bibnamefont {Strange}}, \bibinfo {author} {\bibfnamefont
  {G.~A.}\ \bibnamefont {Tritsaris}}, \bibinfo {author} {\bibfnamefont
  {M.}~\bibnamefont {Vanin}}, \bibinfo {author} {\bibfnamefont
  {M.}~\bibnamefont {Walter}}, \bibinfo {author} {\bibfnamefont
  {B.}~\bibnamefont {Hammer}}, \bibinfo {author} {\bibfnamefont
  {H.}~\bibnamefont {H{\"a}kkinen}}, \bibinfo {author} {\bibfnamefont
  {G.~K.~H.}\ \bibnamefont {Madsen}}, \bibinfo {author} {\bibfnamefont {R.~M.}\
  \bibnamefont {Nieminen}}, \bibinfo {author} {\bibfnamefont {J.~K.}\
  \bibnamefont {N{\o}rskov}}, \bibinfo {author} {\bibfnamefont
  {M.}~\bibnamefont {Puska}}, \bibinfo {author} {\bibfnamefont {T.~T.}\
  \bibnamefont {Rantala}}, \bibinfo {author} {\bibfnamefont {J.}~\bibnamefont
  {Schi{\o}tz}}, \bibinfo {author} {\bibfnamefont {K.~S.}\ \bibnamefont
  {Thygesen}}, \ and\ \bibinfo {author} {\bibfnamefont {K.~W.}\ \bibnamefont
  {Jacobsen}},\ }\href {\doibase 10.1088/0953-8984/22/25/253202} {\bibfield
  {journal} {\bibinfo  {journal} {J. Phys.: Condens. Matter}\ }\textbf
  {\bibinfo {volume} {22}},\ \bibinfo {pages} {253202} (\bibinfo {year}
  {2010})}\BibitemShut {NoStop}%
\bibitem [{GPA()}]{GPAW}%
  \BibitemOpen
  \href {https://wiki.fysik.dtu.dk/gpaw/} {\enquote {\bibinfo {title} {{GPAW}:
  {DFT} and beyond within the projector-augmented wave method},}\ }\bibinfo
  {note} {\url{https://wiki.fysik.dtu.dk/gpaw/}}\BibitemShut {NoStop}%
\bibitem [{\citenamefont {Bl\"ochl}(1994)}]{Blochl1994}%
  \BibitemOpen
  \bibfield  {author} {\bibinfo {author} {\bibfnamefont {P.~E.}\ \bibnamefont
  {Bl\"ochl}},\ }\href {\doibase 10.1103/PhysRevB.50.17953} {\bibfield
  {journal} {\bibinfo  {journal} {Phys. Rev. B}\ }\textbf {\bibinfo {volume}
  {50}},\ \bibinfo {pages} {17953} (\bibinfo {year} {1994})}\BibitemShut
  {NoStop}%
\bibitem [{\citenamefont {Walter}\ \emph {et~al.}(2008)\citenamefont {Walter},
  \citenamefont {H{\"a}kkinen}, \citenamefont {Lehtovaara}, \citenamefont
  {Puska}, \citenamefont {Enkovaara}, \citenamefont {Rostgaard},\ and\
  \citenamefont {Mortensen}}]{Walter2008}%
  \BibitemOpen
  \bibfield  {author} {\bibinfo {author} {\bibfnamefont {M.}~\bibnamefont
  {Walter}}, \bibinfo {author} {\bibfnamefont {H.}~\bibnamefont
  {H{\"a}kkinen}}, \bibinfo {author} {\bibfnamefont {L.}~\bibnamefont
  {Lehtovaara}}, \bibinfo {author} {\bibfnamefont {M.}~\bibnamefont {Puska}},
  \bibinfo {author} {\bibfnamefont {J.}~\bibnamefont {Enkovaara}}, \bibinfo
  {author} {\bibfnamefont {C.}~\bibnamefont {Rostgaard}}, \ and\ \bibinfo
  {author} {\bibfnamefont {J.~J.}\ \bibnamefont {Mortensen}},\ }\href {\doibase
  10.1063/1.2943138} {\bibfield  {journal} {\bibinfo  {journal} {J. Chem.
  Phys.}\ }\textbf {\bibinfo {volume} {128}},\ \bibinfo {pages} {244101}
  (\bibinfo {year} {2008})}\BibitemShut {NoStop}%
\bibitem [{\citenamefont {Wilkinson}(1956)}]{Wilkinson1956}%
  \BibitemOpen
  \bibfield  {author} {\bibinfo {author} {\bibfnamefont {P.~G.}\ \bibnamefont
  {Wilkinson}},\ }\href {\doibase 10.1139/p56-067} {\bibfield  {journal}
  {\bibinfo  {journal} {Can. J. Phys.}\ }\textbf {\bibinfo {volume} {34}},\
  \bibinfo {pages} {596} (\bibinfo {year} {1956})}\BibitemShut {NoStop}%
\bibitem [{\citenamefont {Ferguson}\ \emph {et~al.}(1957)\citenamefont
  {Ferguson}, \citenamefont {Reeves},\ and\ \citenamefont
  {Schneider}}]{Ferguson1957}%
  \BibitemOpen
  \bibfield  {author} {\bibinfo {author} {\bibfnamefont {J.}~\bibnamefont
  {Ferguson}}, \bibinfo {author} {\bibfnamefont {L.~W.}\ \bibnamefont
  {Reeves}}, \ and\ \bibinfo {author} {\bibfnamefont {W.~G.}\ \bibnamefont
  {Schneider}},\ }\href {\doibase 10.1139/v57-152} {\bibfield  {journal}
  {\bibinfo  {journal} {Can. J. Chem.}\ }\textbf {\bibinfo {volume} {35}},\
  \bibinfo {pages} {1117} (\bibinfo {year} {1957})}\BibitemShut {NoStop}%
\bibitem [{\citenamefont {Rossi}\ \emph
  {et~al.}(2015{\natexlab{b}})\citenamefont {Rossi}, \citenamefont {Lehtola},
  \citenamefont {Sakko}, \citenamefont {Puska},\ and\ \citenamefont
  {Nieminen}}]{Rossi2015Nanoplasmonics}%
  \BibitemOpen
  \bibfield  {author} {\bibinfo {author} {\bibfnamefont {T.~P.}\ \bibnamefont
  {Rossi}}, \bibinfo {author} {\bibfnamefont {S.}~\bibnamefont {Lehtola}},
  \bibinfo {author} {\bibfnamefont {A.}~\bibnamefont {Sakko}}, \bibinfo
  {author} {\bibfnamefont {M.~J.}\ \bibnamefont {Puska}}, \ and\ \bibinfo
  {author} {\bibfnamefont {R.~M.}\ \bibnamefont {Nieminen}},\ }\href {\doibase
  10.1063/1.4913739} {\bibfield  {journal} {\bibinfo  {journal} {J. Chem.
  Phys.}\ }\textbf {\bibinfo {volume} {142}},\ \bibinfo {pages} {094114}
  (\bibinfo {year} {2015}{\natexlab{b}})}\BibitemShut {NoStop}%
\bibitem [{\citenamefont {Perdew}\ \emph {et~al.}(1996)\citenamefont {Perdew},
  \citenamefont {Burke},\ and\ \citenamefont {Ernzerhof}}]{Perdew1996}%
  \BibitemOpen
  \bibfield  {author} {\bibinfo {author} {\bibfnamefont {J.~P.}\ \bibnamefont
  {Perdew}}, \bibinfo {author} {\bibfnamefont {K.}~\bibnamefont {Burke}}, \
  and\ \bibinfo {author} {\bibfnamefont {M.}~\bibnamefont {Ernzerhof}},\ }\href
  {\doibase 10.1103/PhysRevLett.77.3865} {\bibfield  {journal} {\bibinfo
  {journal} {Phys. Rev. Lett.}\ }\textbf {\bibinfo {volume} {77}},\ \bibinfo
  {pages} {3865} (\bibinfo {year} {1996})}\BibitemShut {NoStop}%
\bibitem [{\citenamefont {Perdew}\ \emph {et~al.}(1997)\citenamefont {Perdew},
  \citenamefont {Burke},\ and\ \citenamefont {Ernzerhof}}]{Perdew1997}%
  \BibitemOpen
  \bibfield  {author} {\bibinfo {author} {\bibfnamefont {J.~P.}\ \bibnamefont
  {Perdew}}, \bibinfo {author} {\bibfnamefont {K.}~\bibnamefont {Burke}}, \
  and\ \bibinfo {author} {\bibfnamefont {M.}~\bibnamefont {Ernzerhof}},\ }\href
  {\doibase 10.1103/PhysRevLett.78.1396} {\bibfield  {journal} {\bibinfo
  {journal} {Phys. Rev. Lett.}\ }\textbf {\bibinfo {volume} {78}},\ \bibinfo
  {pages} {1396} (\bibinfo {year} {1997})}\BibitemShut {NoStop}%
\bibitem [{Not({\natexlab{a}})}]{NoteTaito}%
  \BibitemOpen
  \bibinfo {note} {The timing was performed on the Taito supercluster of CSC --
  IT Center for Science, Finland. Each computing node has Intel Haswell
  E5-2690v3 processors and the nodes are connected with Infiniband FDR
  interconnect. For further details of the hardware, see
  \url{https://research.csc.fi/taito-supercluster} (9th Feb 2017).}\BibitemShut
  {Stop}%
\bibitem [{\citenamefont {Gritsenko}\ \emph {et~al.}(1995)\citenamefont
  {Gritsenko}, \citenamefont {van Leeuwen}, \citenamefont {van Lenthe},\ and\
  \citenamefont {Baerends}}]{Gritsenko1995}%
  \BibitemOpen
  \bibfield  {author} {\bibinfo {author} {\bibfnamefont {O.}~\bibnamefont
  {Gritsenko}}, \bibinfo {author} {\bibfnamefont {R.}~\bibnamefont {van
  Leeuwen}}, \bibinfo {author} {\bibfnamefont {E.}~\bibnamefont {van Lenthe}},
  \ and\ \bibinfo {author} {\bibfnamefont {E.~J.}\ \bibnamefont {Baerends}},\
  }\href {\doibase 10.1103/PhysRevA.51.1944} {\bibfield  {journal} {\bibinfo
  {journal} {Phys. Rev. A}\ }\textbf {\bibinfo {volume} {51}},\ \bibinfo
  {pages} {1944} (\bibinfo {year} {1995})}\BibitemShut {NoStop}%
\bibitem [{\citenamefont {Kuisma}\ \emph {et~al.}(2010)\citenamefont {Kuisma},
  \citenamefont {Ojanen}, \citenamefont {Enkovaara},\ and\ \citenamefont
  {Rantala}}]{Kuisma2010}%
  \BibitemOpen
  \bibfield  {author} {\bibinfo {author} {\bibfnamefont {M.}~\bibnamefont
  {Kuisma}}, \bibinfo {author} {\bibfnamefont {J.}~\bibnamefont {Ojanen}},
  \bibinfo {author} {\bibfnamefont {J.}~\bibnamefont {Enkovaara}}, \ and\
  \bibinfo {author} {\bibfnamefont {T.~T.}\ \bibnamefont {Rantala}},\ }\href
  {\doibase 10.1103/PhysRevB.82.115106} {\bibfield  {journal} {\bibinfo
  {journal} {Phys. Rev. B}\ }\textbf {\bibinfo {volume} {82}},\ \bibinfo
  {pages} {115106} (\bibinfo {year} {2010})}\BibitemShut {NoStop}%
\bibitem [{\citenamefont {Yan}\ \emph {et~al.}(2011)\citenamefont {Yan},
  \citenamefont {Jacobsen},\ and\ \citenamefont {Thygesen}}]{Yan2011First}%
  \BibitemOpen
  \bibfield  {author} {\bibinfo {author} {\bibfnamefont {J.}~\bibnamefont
  {Yan}}, \bibinfo {author} {\bibfnamefont {K.~W.}\ \bibnamefont {Jacobsen}}, \
  and\ \bibinfo {author} {\bibfnamefont {K.~S.}\ \bibnamefont {Thygesen}},\
  }\href {\doibase 10.1103/PhysRevB.84.235430} {\bibfield  {journal} {\bibinfo
  {journal} {Phys. Rev. B}\ }\textbf {\bibinfo {volume} {84}},\ \bibinfo
  {pages} {235430} (\bibinfo {year} {2011})}\BibitemShut {NoStop}%
\bibitem [{\citenamefont {Yan}\ \emph {et~al.}(2012)\citenamefont {Yan},
  \citenamefont {Jacobsen},\ and\ \citenamefont {Thygesen}}]{Yan2012}%
  \BibitemOpen
  \bibfield  {author} {\bibinfo {author} {\bibfnamefont {J.}~\bibnamefont
  {Yan}}, \bibinfo {author} {\bibfnamefont {K.~W.}\ \bibnamefont {Jacobsen}}, \
  and\ \bibinfo {author} {\bibfnamefont {K.~S.}\ \bibnamefont {Thygesen}},\
  }\href {\doibase 10.1103/PhysRevB.86.241404} {\bibfield  {journal} {\bibinfo
  {journal} {Phys. Rev. B}\ }\textbf {\bibinfo {volume} {86}},\ \bibinfo
  {pages} {241404} (\bibinfo {year} {2012})}\BibitemShut {NoStop}%
\bibitem [{\citenamefont {He}\ and\ \citenamefont {Zeng}(2010)}]{He2010}%
  \BibitemOpen
  \bibfield  {author} {\bibinfo {author} {\bibfnamefont {Y.}~\bibnamefont
  {He}}\ and\ \bibinfo {author} {\bibfnamefont {T.}~\bibnamefont {Zeng}},\
  }\href {\doibase 10.1021/jp101598j} {\bibfield  {journal} {\bibinfo
  {journal} {J. Phys. Chem. C}\ }\textbf {\bibinfo {volume} {114}},\ \bibinfo
  {pages} {18023} (\bibinfo {year} {2010})}\BibitemShut {NoStop}%
\bibitem [{\citenamefont {Krauter}\ \emph {et~al.}(2015)\citenamefont
  {Krauter}, \citenamefont {Bernadotte}, \citenamefont {Jacob}, \citenamefont
  {Pernpointner},\ and\ \citenamefont {Dreuw}}]{Krauter2015}%
  \BibitemOpen
  \bibfield  {author} {\bibinfo {author} {\bibfnamefont {C.~M.}\ \bibnamefont
  {Krauter}}, \bibinfo {author} {\bibfnamefont {S.}~\bibnamefont {Bernadotte}},
  \bibinfo {author} {\bibfnamefont {C.~R.}\ \bibnamefont {Jacob}}, \bibinfo
  {author} {\bibfnamefont {M.}~\bibnamefont {Pernpointner}}, \ and\ \bibinfo
  {author} {\bibfnamefont {A.}~\bibnamefont {Dreuw}},\ }\href {\doibase
  10.1021/acs.jpcc.5b07659} {\bibfield  {journal} {\bibinfo  {journal} {J.
  Phys. Chem. C}\ }\textbf {\bibinfo {volume} {119}},\ \bibinfo {pages} {24564}
  (\bibinfo {year} {2015})}\BibitemShut {NoStop}%
\bibitem [{Not({\natexlab{b}})}]{NoteSupplement}%
  \BibitemOpen
  \bibinfo {note} {See Supplemental Material for additional transition
  contribution maps for the Ag$_{55}$ nanoparticle.}\BibitemShut {Stop}%
\bibitem [{\citenamefont {Bae}\ and\ \citenamefont {Aikens}(2012)}]{Bae2012}%
  \BibitemOpen
  \bibfield  {author} {\bibinfo {author} {\bibfnamefont {G.-T.}\ \bibnamefont
  {Bae}}\ and\ \bibinfo {author} {\bibfnamefont {C.~M.}\ \bibnamefont
  {Aikens}},\ }\href {\doibase 10.1021/jp300789x} {\bibfield  {journal}
  {\bibinfo  {journal} {J. Phys. Chem. C}\ }\textbf {\bibinfo {volume} {116}},\
  \bibinfo {pages} {10356} (\bibinfo {year} {2012})}\BibitemShut {NoStop}%
\bibitem [{\citenamefont {Rabilloud}(2014)}]{Rabilloud2014}%
  \BibitemOpen
  \bibfield  {author} {\bibinfo {author} {\bibfnamefont {F.}~\bibnamefont
  {Rabilloud}},\ }\href {\doibase 10.1063/1.4897260} {\bibfield  {journal}
  {\bibinfo  {journal} {J. Chem. Phys.}\ }\textbf {\bibinfo {volume} {141}},\
  \bibinfo {pages} {144302} (\bibinfo {year} {2014})}\BibitemShut {NoStop}%
\end{thebibliography}%

\end{document}